\documentclass[aip,pof,amsmath,amssymb]{revtex4-1}

\usepackage{graphicx, epstopdf}
\usepackage{dcolumn}
\usepackage{color}
\usepackage{bm}
\usepackage[colorlinks=true,       
    linkcolor=blue,          
    citecolor=blue,
    urlcolor=blue]{hyperref}
\usepackage{multirow}

\begin{document}


\title[Precessing annulus]{Experimental study of fluid flows in a precessing cylindrical annulus}
\author{Yufeng Lin}
\email{yufeng.lin@erdw.ethz.ch}
\affiliation{Institute of Geophysics, ETH Z\"urich, Sonneggstrasse 5, 8092 Z\"urich, Switzerland }
\author{Jerome Noir}
\author{Andrew Jackson}

\date{\today}

\begin{abstract}
The flow inside a precessing fluid cavity has been given particular attention since the end of the 19th century in geophysical and industrial contexts.
 The present study aims at shedding light on the underlying mechanism by which the flow inside a precessing cylindrical annulus transitions from laminar to multiple scale complex structures. We address this problem experimentally using ultrasonic Doppler velocimetry to diagnose the fluid velocity in a rotating and precessing cylindrical annulus. When precession is weak, the flow can be described as a superposition of forced inertial modes. Above a critical value of the precession rate, the forced flow couples with two free inertial modes satisfying triadic resonance conditions, leading to the classical growth and collapse. Using a  Bayesian approach, we extract the wavenumber, frequency, growth rate and amplitude of each mode involved in the instability. In some cases, we observe for the first time ever experimentally two pairs of free modes coexisting with the forced flow. At larger precession rates, we do not observe triadic resonance any more, instead we observe several harmonics whose frequencies are  integer multiples of the rotation frequency. 
\end{abstract}


\maketitle

\section{Introduction \label{Intr}}

The problem of precession and nutation driven flows has been studied since the end of the 19th century in geophysical/astrophysical and industrial contexts. In the planetary community, precession and nutation appeared as a proxy to probe the internal structure of planets\cite{Poincare1910} and as a possible energy source for dynamo processes\cite{Malkus1968}. In the aerospace industry, precession driven flows play an important role in the question of the stability of spacecraft with liquid payloads.

In a sphere and a weakly deformed spheroid, precession induces a flow of uniform vorticity,  the so-called Poincar\'e mode, which is a quasi solid body rotation around an axis tilted from the rotation axis of the container\cite{Poincare1910}. In addition, oblique and geostrophic shear layers are spawned from the boundary layer at critical latitudes \cite{ Stewartson1963,Busse1968,Kerswell1995, Kida2011}, which have been confirmed by numerical\cite{Hollerbach1995,Noir2001} and experimental\cite{Malkus1968,Vanyo1995,Noir2001b,Noir2003}studies.  
The uniform vorticity solution was also extended to precessing triaxial ellipsoids recently\cite{Noir2013}. The case of a spherical shell, i.e. with a solid inner core at the center of the fluid, has been investigated numerically \cite{Hollerbach1995,Tilgner1999a,Tilgner2001} and experimentally\cite{Triana2012} showing little influence on the first order flow, but with additional internal shear layers being spawned from the inner boundary layer. Theoretical investigation is not possible in this geometry due to the ill-posed Cauchy problem inherent to the spherical shell\cite{Stewartson1969}. At large enough precession rate, the uniform vorticity flow becomes unstable through the parametric coupling of two inertial waves leading to a growth and collapse to small scales \cite{Kerswell1993,Lorenzani2001,Lorenzani2003,Wei2013}. 

The flow inside a cylinder subject to precession has been investigated experimentally\cite{Manasseh1992,Kobine1995,Kobine1996,Meunier2008,Lagrange2008, Lagrange2011,Mouhali2012}, numerically\cite{Liao2012} and theoretically\cite{Gans1970,Mahalov1993,Meunier2008,Lehner2010,Liao2012}. The cylindrical geometry differs significantly from the spherical or spheroidal one as the forced flow does not necessarily reduce to a uniform vorticity solution or Poincar\'e mode. Depending on the aspect ratio of the cavity, different inertial modes could be in resonance in a precessing cylinder\cite{McEwan1970,Manasseh1992,Kobine1995,Meunier2008,Liao2012}. At resonance, whereby the frequency of an inertial mode with azimuthal wavenumber $m=1$ is equal to the frequency of the precessional force, the flow can be approximated as a single resonant inertial mode, and away from the resonance, the flow can be described as a superimposition of inertial modes\cite{Gans1970,Liao2012}. Previous experiments have shown that precessionally forced flow in a cylinder can be unstable and break down to small scale disorder\cite{McEwan1970,Manasseh1992}. Lagrange et al.\cite{Lagrange2011} have shown that the precessional instability in a cylinder can be explained by the mechanism of triadic resonance, i.e. the forced inertial mode can excite two free inertial modes via parametric coupling if the difference in the wavenumber and frequency between the two free modes matches the wavenumber and frequency of the forced mode. The triadic resonance has been confirmed by laboratory experiments using particle image velocimetry (PIV) measurements\cite{Lagrange2008,Lagrange2011}. As in a spheroidal container, the instability is characterized by  distinct growth and collapse of the free modes. The growth phase is well explained by the triadic resonance, however, the underlying mechanism of the collapse phase is still not well established.   
Finally, numerical studies have shown that the flow driven by precession can sustain dynamos in a sphere\cite{Tilgner2005,Tilgner2007a}, spheroid\cite{Wu2009} and cylinder\cite{Nore2011}. 

In the present study, we investigate fluid flows driven by precession in a cylindrical annulus of moderate radius ratio via laboratory experiments. Unlike a spherical shell, the linear inviscid eigen-value problem describing the inertial mode is mathematically tractable in a cylindrical annulus\cite{Zhang2010b}. Hence, a cylindrical annulus can be seen as the prototype model to study the effect of a solid inner core in planetary liquid bodies because of its mathematical simplicity. Motivated by this, we carry out the corresponding experimental study in a rotating and precessing cylindrical annulus filled with water. The fluid flow in the tank is diagnosed by the use of ultrasonic Doppler velocimetry (UDV). We characterize different flow regimes based on the UDV measurements by varying the rotation rate and precession rate over a wide range.  

The remaining part of the paper is organized as following. Section \ref{Math} presents some mathematical formulations, Section \ref{Expe} introduces the experimental setup and experimental results are presented in Section \ref{Rest}. The paper closes with a discussion in Section \ref{Diss}.

\section{Mathematical formulation}\label{Math}
\subsection{Governing equations}
\begin{figure}
  \centerline{\includegraphics[width=8cm]{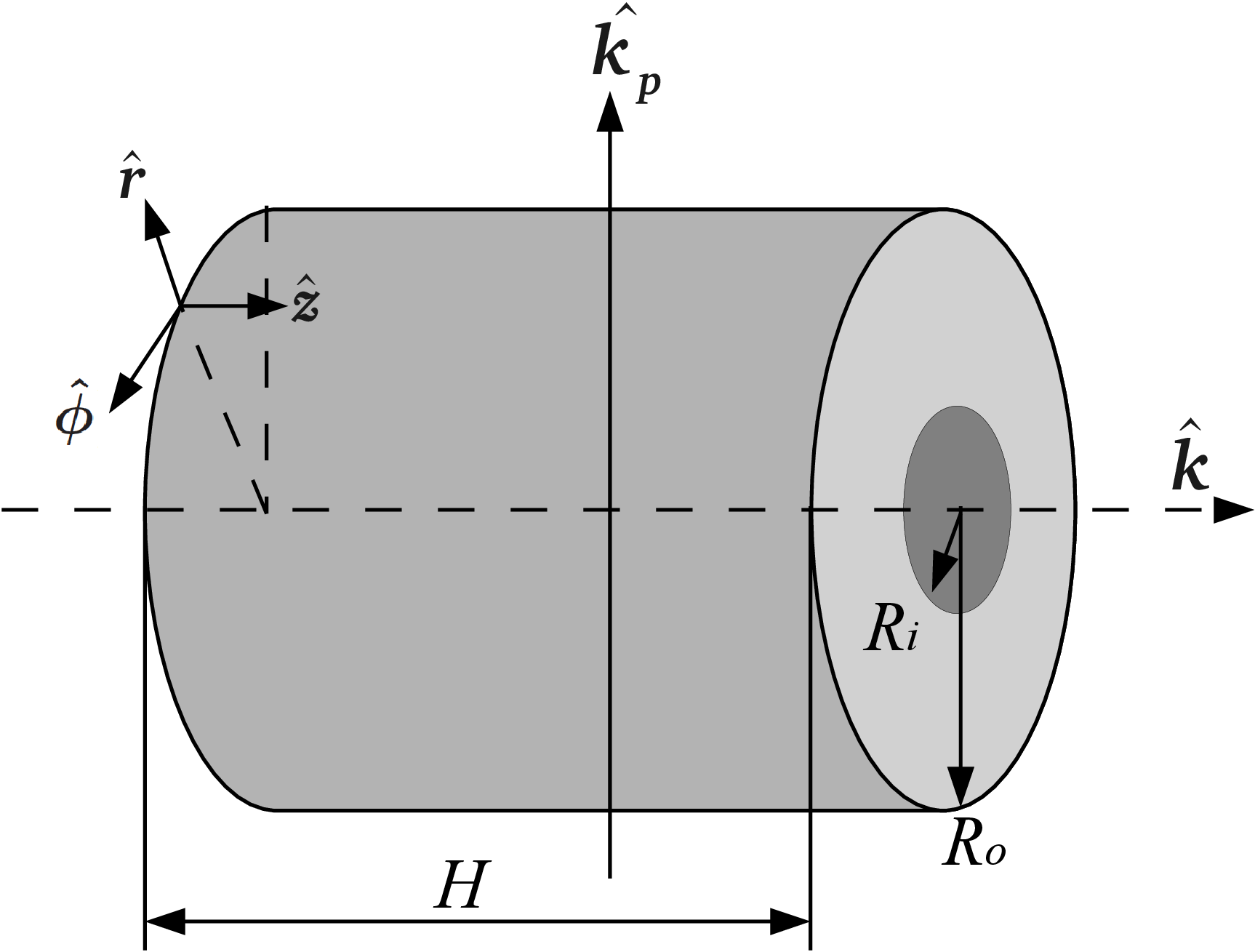}}
  \caption{Sketch of the problem.}
\label{fig:Figure1}
\end{figure}
We consider a cylindrical annulus of inner radius $R_i$, outer radius $R_o$ and height $H$, filled with a homogeneous and incompressible fluid of density $\rho$ and kinematic viscosity $\nu$.  
The cylindrical annulus rotates at $\boldsymbol{\Omega_o}= \Omega_o\bm{\hat k}$ and precesses at $\boldsymbol{\Omega_p}=\Omega_p {\bm{\hat k}}_p$, tilted by $90^{\circ}$ compared to ${\bm{\hat k}}$ (Fig. \ref{fig:Figure1}). In the frame of reference attached to the container, herein referred as the rotating frame, ${\bm{\hat k}}$ is fixed whereas ${\bm{\hat k}}_p$ rotates along ${\bm{\hat k}}$ at $-\Omega_o$. Using a cylindrical coordinate system ($r,\phi, z$) in the rotating frame as sketched in Fig. \ref{fig:Figure1},  the rotation and precession vectors can be expressed as: 
\begin{eqnarray}
   \boldsymbol{\Omega_o} & = & \Omega_o   \boldsymbol {\hat z}, \label{eq:omgo} \\
   \boldsymbol{\Omega_p} & = &  \Omega_p[\cos(\phi+\Omega_o t^*)\boldsymbol {\hat r }-\sin(\phi+\Omega_o t^*)\boldsymbol {\hat \phi}], \label{eq:omgp}
\end{eqnarray}
where ($\boldsymbol {\hat r},\boldsymbol {\hat \phi},\boldsymbol {\hat z}$) are the unit vectors of the cylindrical coordinates and $t^*$ is dimensional time.

Using the outer radius $R_o$ as the length scale and $\Omega_o ^{-1}$ as the time scale so that dimensionless time $t=\Omega_ot^*$, the Navier-Stokes equations governing the fluid velocity $\bm{u}$ can be derived in the rotating frame in a dimensionless form\cite{Meunier2008,Liao2012}
\begin{eqnarray}
   \frac{\partial \boldsymbol u}{\partial t}+\boldsymbol u\cdot \nabla \boldsymbol u +2[\boldsymbol{\hat{z}} +P_o(\cos(\phi+t)\boldsymbol {\hat r }&-& \sin(\phi+t)\boldsymbol {\hat \phi})]\times \boldsymbol u    \nonumber \\ 
  &=&  -\nabla p+E \nabla ^2 \boldsymbol u -2P_o r\cos(\phi+t)\boldsymbol{\hat{z}},   \label{eq:nseq_nd} \\
   \nabla \cdot \boldsymbol u& =&  0 \label{eq:incompressible}, 
\end{eqnarray}
where the Poincar\'e  number $P_o$ and Ekman number $E$ are defined as 
\refstepcounter{equation}
$$
 P_o=\frac{{\Omega_p}}{{\Omega_o}}, \quad 
 E=\frac{\nu}{{\Omega_o}R_o^2}.
 \eqno{(\theequation{\mathit{a},\mathit{b}})}\label{eq:PoEk}
$$
The last term on the left hand side is the Coriolis force due to the rotation and the precession and the last term on the right hand side represents the Poincar\'e force associated with the gyroscopic motion. The dimensionless reduced pressure $p$ includes all the potential terms. 

Finally we introduce aspect ratio $h$ and radius ratio $r_{i}$ 
\refstepcounter{equation}
$$
 h=\frac{H}{R_o}, \quad 
 r_{i}=\frac{R_i}{R_o},
 \eqno{(\theequation{\mathit{a},\mathit{b}})}\label{eq:ratio}
$$
to characterize the geometry of a cylindrical annulus.

\subsection{Inertial mode and triadic resonance}\label{inertialmode}
Oscillatory motions, known as inertial waves or inertial modes, are ubiquitous in a rapidly rotating fluid system where the Coriolis force works as the restoring force. If we neglect viscosity and do not consider any specific excitation force, small perturbations in a uniform rotating fluid are governed by  
\begin{eqnarray}
   \frac{\partial \boldsymbol u}{\partial t}+2\boldsymbol{\hat{z}}\times \boldsymbol u &=&  -\nabla p   \label{eq:nseq_1} \\
   \nabla \cdot \boldsymbol u& =&  0 \label{eq:incompressible_1}.
\end{eqnarray}
Eqs. (\ref{eq:nseq_1}) and (\ref{eq:incompressible_1}) support wavelike solutions  $\bm{u}\propto \mathrm e^{\mathrm{i}(\bm{k}\cdot\bm{r}-\omega t)}$ in an unbounded fluid with the dispersion relation $\omega=\pm 2\bm{k}\cdot\hat{\bm z}/|\bm{k}|$. Here $\bm k$ is the wave vector and we can see that the inertial wave frequency is bounded to $|\omega|\leqslant 2$. In an enclosed fluid, waves reflected on the boundaries lead to constructive interferences, which forms eigen-modes of the cavity, also called inertial modes.
Analytical expressions of inertial modes can be obtained in some simple geometries such as spheres and cylinders. The analytic solution of inertial modes $\bm{u}_{mnk}$ in a cylindrical annulus, which is relevant to this study, is derived in Appendix \ref{App1} where $m,n,k$ are integers representing the azimuthal, axial and radial wavenumber respectively. In the remaining part of the paper, each inertial mode is labelled by its wavenumbers ($m,n,k$) and $\omega_{mnk}$ is the corresponding eigen frequency.      

When an inertial mode ($m_0,n_0,k_0$) is forced at frequency $\omega_0$, it can participate to a parametric resonance with two free inertial modes ($m_1,n_1,k_1$) with frequency $\omega_1$ and ($m_2,n_2,k_2$) with frequency $\omega_2$. The so called triadic resonance can occur if these modes satisfy the following parametric conditions \cite{Kerswell1999} 
\begin{eqnarray}
\omega_2-\omega_1 &= &\omega_0,  \label{eq:con1}\\
m_2 - m_1 & = & m_0, \label{eq:con2} \\
n_2 - n_1 & = &n_0. \label{eq:con3}
\end{eqnarray}
There is no strict constraint on the radial wavenumbers, however, two free modes with $k_1=k_2$ have much larger growth rates than the case $k_1\neq k_2$ (Ref \onlinecite{Eloy2003}).
In an inviscid fluid, these conditions (Eqs. (\ref{eq:con1}-\ref{eq:con3})) are necessary and sufficient. The inviscid growth rate is proportional to the amplitude of the forced mode\cite{Kerswell1999}. For finite viscous fluids, the inviscid growth rate must be sufficiently large to overcome the viscous dissipation which is typically $O(E^{1/2})$ \cite{Kerswell1995a,Zhang2008a}. Meanwhile, the viscosity also introduces a viscous detuning, a broadening of the inviscid frequency of order $O(E^{1/2})$ , which facilitates the parametric coupling between modes only satisfying $\omega_2-\omega_1-\omega_0 \approx O(E^{1/2})$. Hence, the selection of the free modes depends on which of the damping or detuning effects is dominant. More detailed derivation and discussion concerning the triadic resonance of inertial modes can be found in Refs. \onlinecite{Kerswell1999, Lagrange2011}. The triadic resonance is an efficient process to transfer and store energy at frequencies different from that of the forcing.    
 
\section{Experimental setup}\label{Expe}
\begin{figure}
  \centerline{\includegraphics[width=9cm]{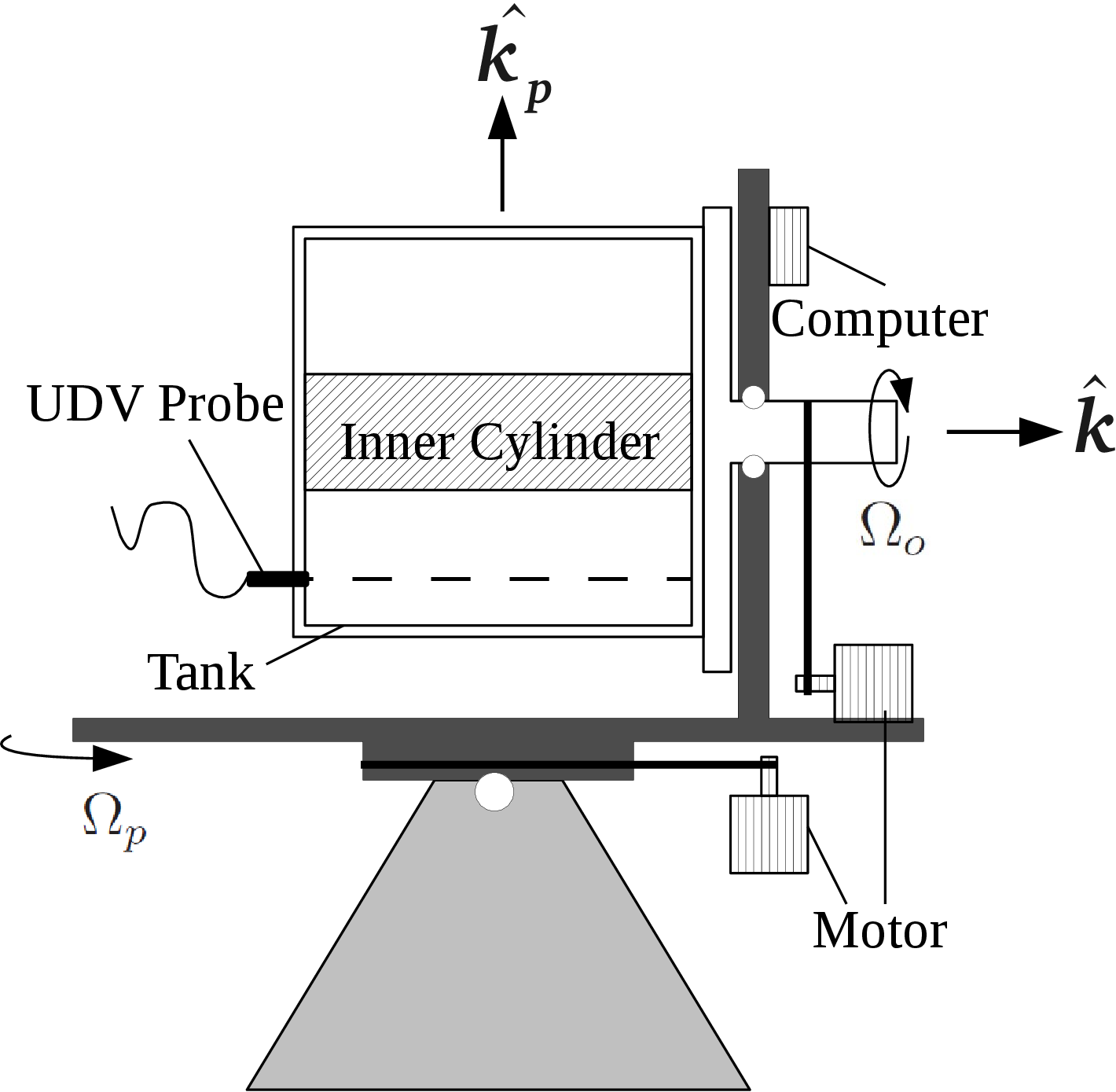}}
  \caption{Sketch of the experimental setup.}
\label{fig:Figure2}
\end{figure}
 
Fig. \ref{fig:Figure2} shows the experimental setup used in this study. The annulus channel is formed of an outer acrylic cylinder of radius and height $R_o=H=142.5$mm and an inner white opaque plastic coaxial cylinder of radius $R_i=38.3$mm. The container is completely filled with water. A variable frequency drive with a timing belt is used to spin the container about its axis of symmetry at 1 rad s$^{-1}\leqslant{\Omega}_o\leqslant$ 20 rad s$^{-1}$. The rotating container sits on a turntable, which rotates at 0.018 rad s$^{-1}\leqslant{\Omega}_p\leqslant$ 1 rad s$^{-1}$,  driven by a second variable frequency drive through a v-belt. Both drives are computer controlled to ensure high accuracy ($\sim 0.1\%$) of the rotation and the precession rate.  
\begin{table}
\caption{Physical and dimensionless parameter definitions and their typical values in this experimental study.}
  \label{tab:parameters}
 \begin{ruledtabular}
  \begin{tabular}{lll}
   Parameter  \ \   & Definition \ \ \ \ \ \ \ \ \     & Experiment \\  [3pt]  
   \hline
$R_o$ &  Outer radius & 142.5 mm  \\ 
$R_i$ &  Inner radius & 38.3 mm  \\ 
$H$ &  Height & 142.5 mm  \\ 
$\Omega_o$ &  Rotation rate \ \ \  & 1-20 rad s$^{-1}$   \\
$\Omega_p$ &  Precession rate \ \ \ \ \ \   & 0.018-1 rad s$^{-1}$   \\
$\nu$ &  Kinematic viscosity (20 $^{\circ}$C)\ \ \  & $10^{-6}$m $^2$ s$^{-1}$   \\ [5pt]

$h$ & Aspect ratio $H/R_o$ & 1.0 \\
$r_{i}$ & Radius ratio $R_i/R_o$ & 0.269 \\
$E$ & Ekman number $\nu/(\Omega_o R_o^2)$ \ \ \ &$\sim2.5\times  10^{-6}$ - $ 5.0\times 10^{-5}$ \\
$P_o$ & Poincar\'e number $\Omega_p/\Omega_o$ \ \ \ &$ \sim $ $10^{-3}$ -1.0 \\
      \end{tabular}
  \end{ruledtabular}
\end{table}

The flow in the container is diagnosed using an UDV (DOP3010 by Signal Processing SA). The UDV measures fluid velocities by sending pulsed ultrasonic waves and detecting the Doppler shift of the reflected signals from particles suspended in the fluid (see http://www.signal-processing.com for more details). Using this technique, we measure the velocity component in the direction of the emitted beam at several locations almost simultaneously. In our experiments, the UDV probe is attached to the end wall of the outer cylinder at a radius $r=120$ mm, providing time resolved profiles along the $z$-direction of the $z$-component of velocity. The typical sampling frequency is around 10 Hz. 

The typical values of parameters explored in this experimental study are summarized in Table \ref{tab:parameters}. Each individual experiment follows the same protocol, we set the container in rotation at ${\Omega_o}$ and wait until the fluid is in rigid rotation with the container, which takes typically about 10 minutes. Then we start precessing the container quickly after we start measuring the velocity. 

\section{Results}\label{Rest}
\subsection{Stable flow} \label{sec:stable}
\begin{figure}
 \includegraphics[width=15cm]{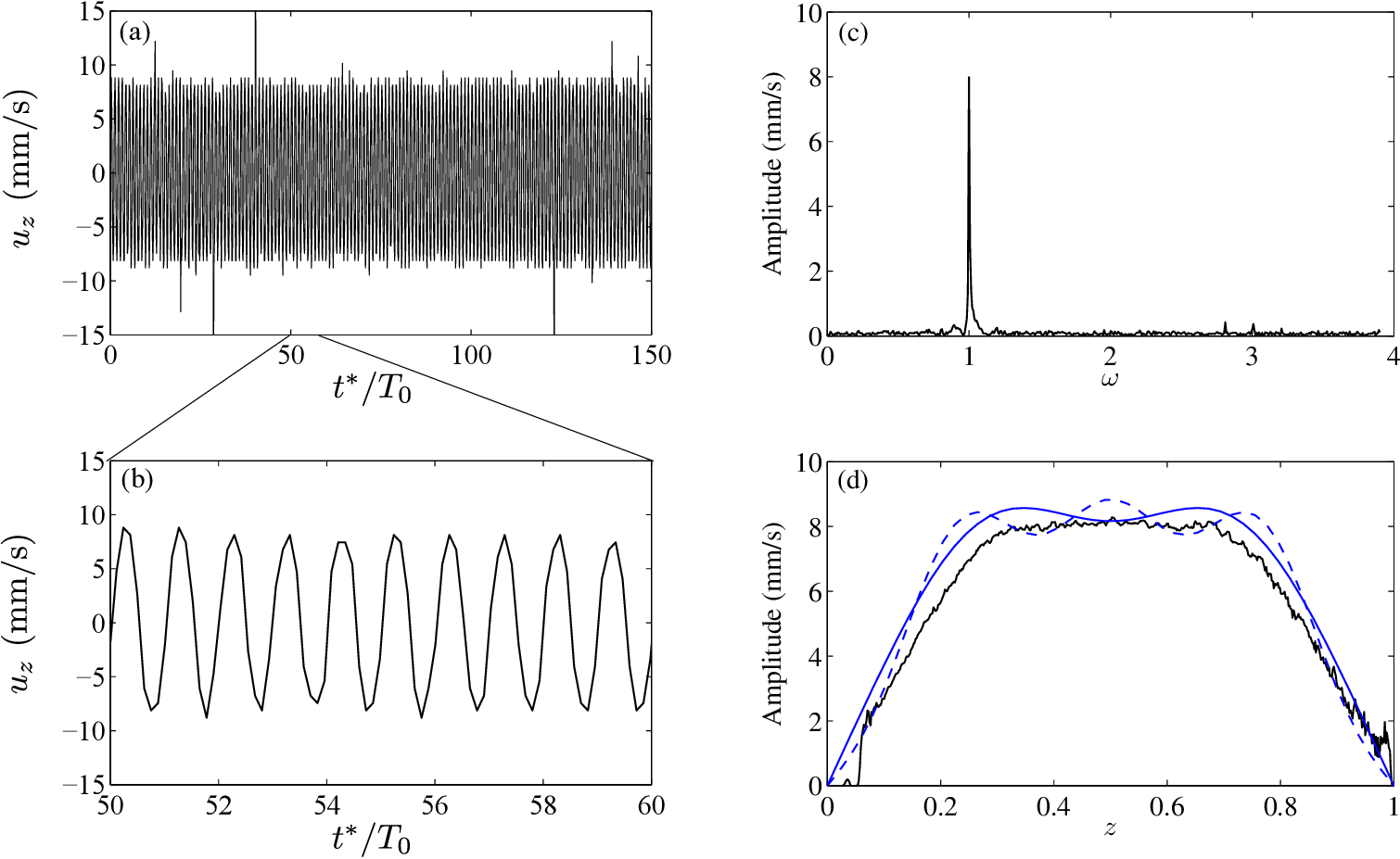}
 \caption{(a) Time series of the axial velocity at $z=0.35$ recorded by the UDV at $P_o=7.0\times10^{-3}$, $E=1.0\times10^{-5}$. Dimensional time $t^*$ has been normalized by $T_0=2\pi/\Omega_o$ which is the rotation period of the tank (same as in other figures). (b) Zoom in of (a) between $t^*/T_0=50$ and $t^*/T_0=60$. (c) Amplitude of the DFT of the time series (a) as a function of frequency $\omega$, the frequency has been normalized by $\Omega_o$ (same as in other figures). (d) Amplitude of the DFT at $\omega=1.0$ as a function of profile depth $z$. Black line represents the experimental data and {blue lines represent the linear inviscid theory given in Appendix \ref{App2} with the truncation $m=1$, $n\leqslant 5$, $k\leqslant 5$ (blue solid line) and $m=1$, $n\leqslant 10$, $k\leqslant 10$ (blue dashed line).}}
\label{fig:Figure3}
\end{figure}

At weak precession, depending on the geometry of the container, the stable flow can be described as a single inertial mode with $m=1$ at resonance, or a linear combination of $m=1$ inertial modes  when far from resonance \cite{Liao2012}. The analytical formulations of inertial modes in a cylindrical annulus are given in Appendix  \ref{App1} and the amplitude of each mode forced by precession is given in Appendix \ref{App2}. With an aspect ratio $h=1.0$ and radius ratio $r_{i}=0.269$ used in this study, there is no resonance ($|\omega_{1nk}-1.0|>E^{1/2}$) with lower order ($n\leqslant5$, $k\leqslant5$) inertial modes provided $E<10^{-4}$ (see Table \ref{tab:InviscidAmp}). 
Fig. \ref{fig:Figure3} shows an example of UDV measurements for such a stable flow at $P_o=7.0\times10^{-3}$ and $E=1.0\times10^{-5}$. The axial velocities recorded by the UDV at depth $z=0.35$ are plotted as a function of time in Fig. \ref{fig:Figure3} (a) and (b). We can see an oscillatory motion with constant amplitude. As shown in Fig. \ref{fig:Figure3} (c), the Discrete Fourier Transform (DFT) exhibits only one well identified dimensionless frequency $\omega=1.0$ which is exactly the frequency of the Poincar\'e force. 
In order to show the spatial structure, we perform the same spectral analysis at all measurement points along $z$ and extract the amplitude at $\omega=1.0$ (Fig. \ref{fig:Figure3} (d)). The black line represents the experimental result and {blue lines represent the linear inviscid theory 
given in Appendix \ref{App2} with the truncation $m=1$, $n\leqslant5$, $k\leqslant5$ (blue solid line) and $m=1$, $n\leqslant10$, $k\leqslant10$ (blue dashed line). Although there are some little discrepancies among them due to the limitation of the inviscid approximation \citep{Liao2012}, both experimental result and linear inviscid solutions with small truncation parameters show that the forced flow is dominated by inertial modes with axial wavenumber $n=1$. } 

\subsection{Triadic resonance } \label{sec:triadic}

\begin{figure}
\includegraphics[width=16cm]{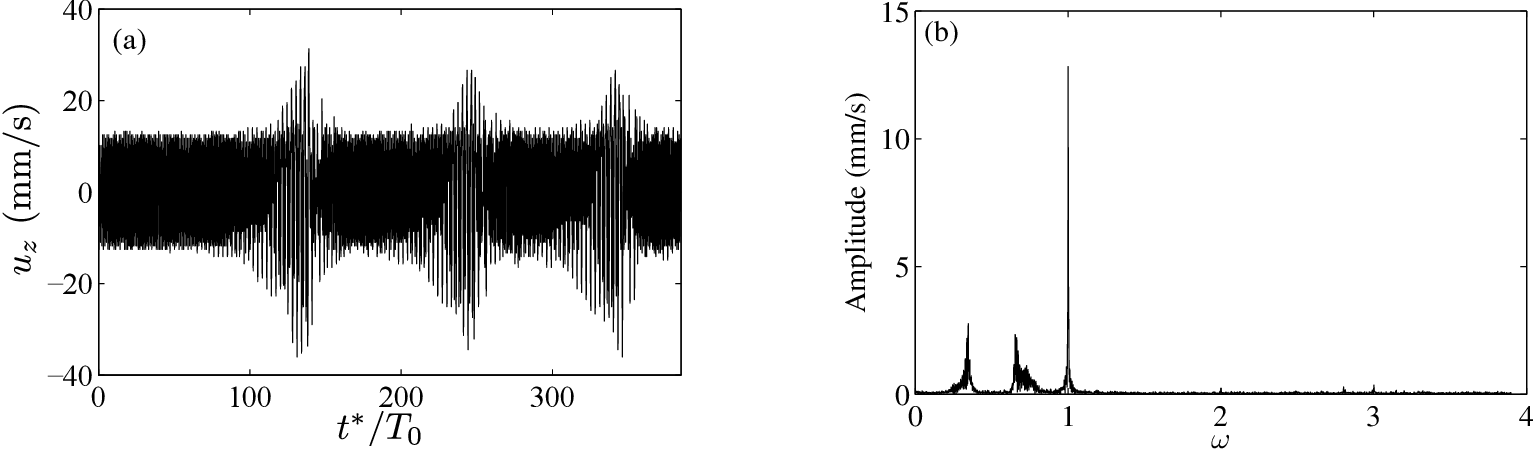}
  \caption{(a) Time series of the axial velocity at $z=0.35$ recorded by the UDV at $P_o=1.0\times10^{-2}$, $E=1.0\times10^{-5}$. (b) Amplitude of the DFT of the time series.}
\label{fig:Figure4}
\end{figure}

\begin{figure}
  \includegraphics[width=12cm]{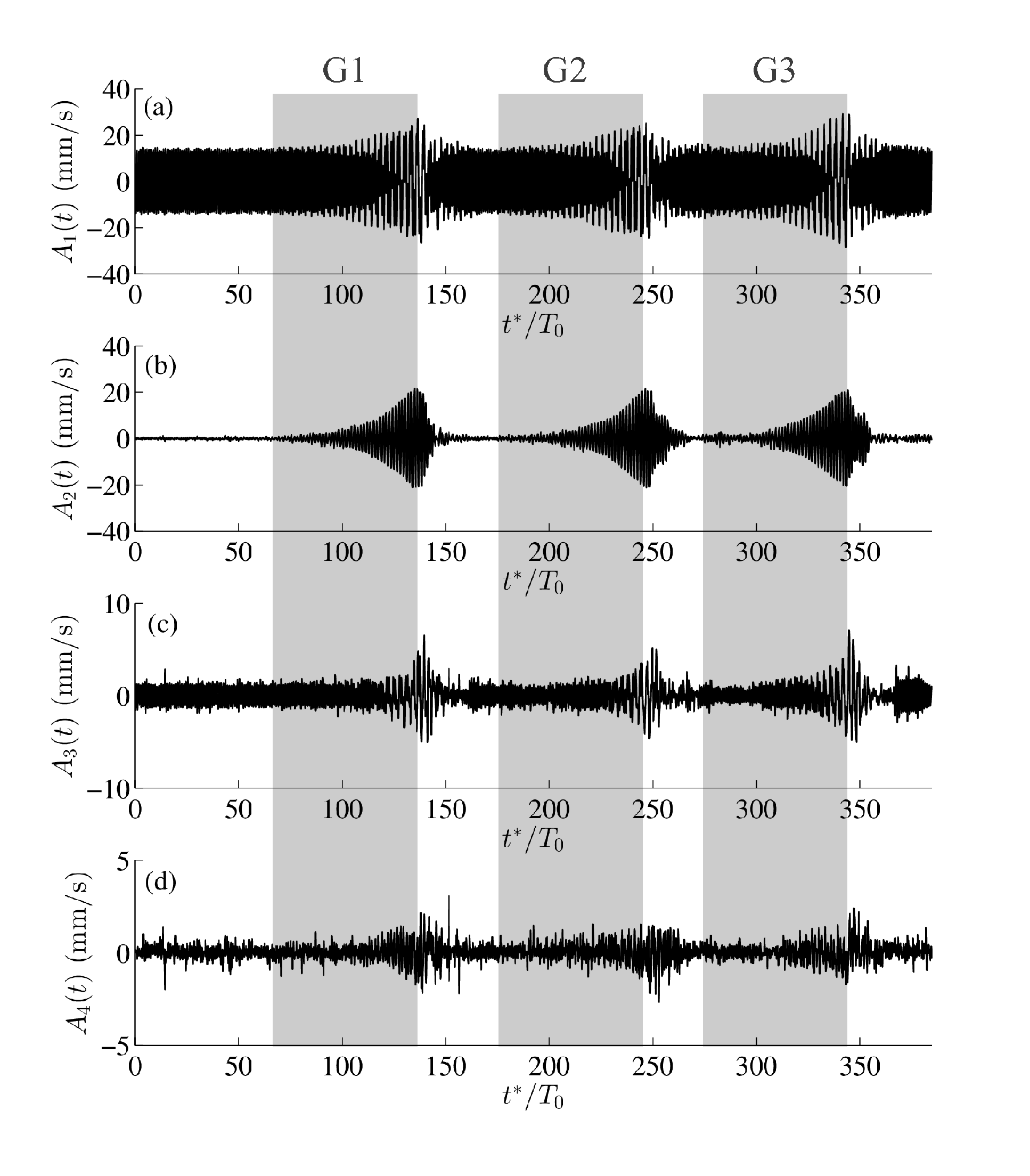}
  \caption{Amplitudes $A_{n}(t)$ at $P_o=1.0\times10^{-2}$, $E=1.0\times10^{-5}$. (a) $n=1$, (b) $n=2$, (c) $n=3$, (d) $n=4$. The grey bars represent time windows G1, G2, G3 used for growth phase analysis.}
\label{fig:Figure5}
\end{figure}

{ As we shall discover in Section \ref{sec:diagram}}, above a critical value of the Poincar\'e number the flow becomes unstable at a given Ekman number. Fig. \ref{fig:Figure4} shows the time series of the axial velocity at $z=0.35$ recorded by the UDV and the corresponding frequency spectrum at $P_o=1.0\times10^{-2}$ and $E=1.0\times10^{-5}$. We observe quiet periods where the signal is similar to the low $P_o$ regime. In addition, we observe growth and collapse phases that are characteristic of the triadic resonances between the forced flow and two free inertial modes \cite{Kerswell1999}. Spectral analysis  using the DFT shows three 
frequency peaks in this moderate $P_o$ regime. Besides the frequency peak at $\omega_0=1.0$ which represents the forced flow, two dominant frequency peaks at $\omega_1=0.66$ and $\omega_2=0.34$ are observed. Note that our UDV measurements can not distinguish positive (retrograde) or negative (prograde) frequencies, so the observed frequency is actually the absolute value of the frequency. In order to compare with experimental observations, Eq. (\ref{eq:con1}) can be rewritten as $|\omega_2|\pm|\omega_1|=1.0$. Obviously, the observed frequencies satisfy the triadic resonance condition.
Assuming a triadic resonance mechanism, the wavenumbers should also satisfy the triadic resonance conditions and the growth rates of the two free modes should be identical. 

We take advantage of the spatial information from the axial velocity profiles obtained by the UDV measurements. Assuming that inertial modes constitute a complete basis (the completeness of inertial modes has been mathematically proven in a cylindrical annulus with very large inner cylinder\cite{Cui2013}), any flow can be represented as a sum of inertial modes \cite{Greenspan1968,Liao2012}. Therefore, the UDV measurement of the axial velocity  $u_z(z,t)$ at a given time can be represented as a sum of sinusoidal functions in $z$ (see Eq. \ref{eq:AppUz} in Appendix \ref{App1} for the axial structure of inertial modes) 
\begin{equation}
u_z(z,t)=\sum_{n=1}^{N} A_{n}(t)\sin( n\pi z/h),
\label{eq:time_wavenumber}
\end{equation}
and then $A_n(t)$ for each axial wavenumber $n$ is given as
\begin{equation}
A_{n}(t)=\frac{\int_0^{h} u(z,t)\sin( n\pi z/h)\, \mathrm {d}z}{\int_0^{h} \sin( n\pi z/h)\sin( n\pi z/h)\,\mathrm{d}z}=2\int_0^{h} u(z,t)\sin( n\pi z/h)\, \mathrm {d}z.
\end{equation}

Fig. \ref{fig:Figure5} shows an example of such a decomposition at $P_o=1.0\times10^{-2}$ and $E=1.0\times10^{-5}$. 
The growth and collapse phases seem to occur on two different time scales, the former being longer than the latter, which suggest they are associated with two different mechanisms and as such should be regarded separately. 
\subsubsection{Growth phase}
\begin{figure}
\includegraphics[width=12cm]{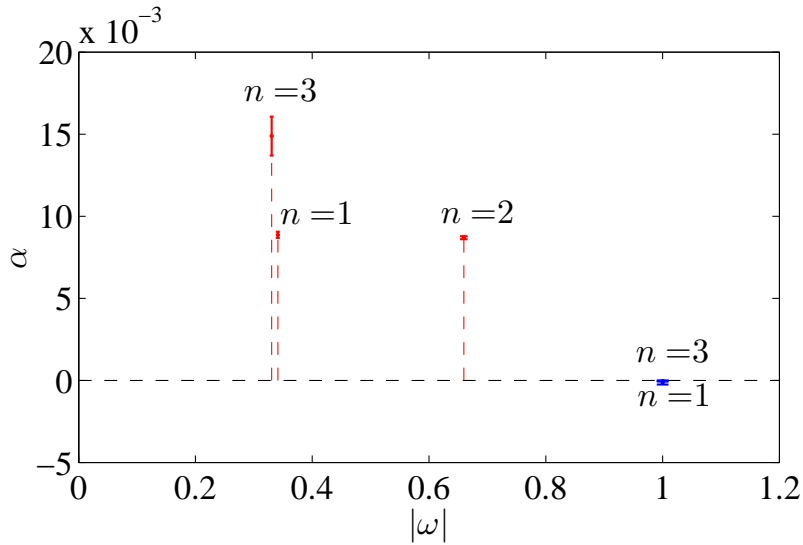}
\caption{Growth rate $\alpha$ as a function of frequency $|\omega|$ obtained from the BPE for the first growth phase G1 at $P_o=1.0\times10^{-2}$,  $E=1.0\times10^{-5}$. The frequencies and growth rates are normalized by the rotation frequency $\Omega_o$. The length of the error bars indicates uncertainties of the growth rates $\alpha$. The blue lines represent the forced flow and the red lines represent the free modes.}
\label{fig:Figure6}
\end{figure}
\begin{table}
\caption{Results of the BPE for the first growth phase G1 at $P_o=1.0\times10^{-2}$,  $E=1.0\times10^{-5}$, and possible inviscid free modes matching the experimental observations. The frequencies and growth rates are normalized by the rotation frequency $\Omega_o$ and uncertainties are given as the standard deviation (same as other BPE results).}
  \begin{ruledtabular}
  \begin{tabular}{lrrrc}
      $n$ & Frequency ($|\omega|$)\ \ \ & Growth rate ($\alpha$)\ \ \ \    &  Amplitude ($B$ (mm/s)) &  \\  [3pt]      
\hline
       1 & $1.0000(\pm3.0\times 10^{-4})$ & $-4.4\times 10^{-5}(\pm 3.0\times 10^{-5})$ & $14.26(\pm6.7\times 10^{-2})$ &  Forced flow \\
       1 & $0.3410(\pm2.0\times 10^{-4})$ & $8.9\times 10^{-3}(\pm 1.9\times 10^{-4})$ & $1.75(\pm6.0\times 10^{-2})$ & Free mode  $\omega_{214}=-0.3363$ \\
       2 & $0.6596(\pm3.4\times 10^{-4})$ & $8.7 \times 10^{-3}(\pm 1.0\times 10^{-4})$ & $2.21(\pm1.2\times 10^{-3})$ & Free mode  $\omega_{324}=0.6862$ \\
       3 & $1.0000(\pm1.3\times 10^{-4})$ & $-1.3 \times 10^{-4}(\pm 1.2\times 10^{-4})$ & $1.29(\pm2.2\times 10^{-2})$ & Forced flow \\
       3 & $0.3305(\pm1.8\times 10^{-3})$ & $1.5 \times 10^{-2}(\pm 1.2\times 10^{-3})$ & $0.04(\pm2.5\times 10^{-2})$ & ? \\
  \end{tabular}
  \label{tab:growth1_R50P03}
  \end{ruledtabular}
\end{table}
During each growth phase G1, G2, G3 (see Fig. \ref{fig:Figure5}), we assume that the time series $A_n(t)$ corresponding to the axial wavenumber $n$ can be modeled as a sum of several harmonic modes with exponentially growing amplitudes,
\begin{equation}
A_n(t)=\sum_{j=1} ^{J}B_{nj}\mathrm e^{\alpha_{nj} t}\cos(\omega_{nj} t+\varphi_{nj}).
\label{eq:growth_model}
\end{equation}

Using the Bayesian parameter estimation (BPE) methodology (Ref \onlinecite{Bretthorst1988,Gregory2010}), 
we extract the frequencies, growth rates and amplitudes of the dominant components corresponding to the axial wavenumber $n$. {The BPE fits a parametric model of the form Eq. (\ref{eq:growth_model}) to the data by working in the time domain; the Fourier transform is avoided as it would be a suboptimal
estimation procedure for a signal that is not stationary, as is the case for our growing or decaying signals. Additionally, there is no need for windowing of the data series to avoid end effects.} 
Within each subset of time series G1, G2, G3, we define a local time with $t=0$ in the middle of the subset time window for the BPE, so $B_{nj}$ represents the amplitude at the middle of the time window. We do not know how many components should be included in the Eq. (\ref{eq:growth_model}) for the BPE, so we start from $J=1$ and increase $J$ until the data are well recovered by the model (Correlation $\geqslant0.95$).  

The results of the BPE for the first growth phase G1 at $P_o=1.0\times10^{-2}$ and $E=1.0\times10^{-5}$ are plotted in Fig. \ref{fig:Figure6} and summarized in Table \ref{tab:growth1_R50P03}. Only the results for $n\leqslant3$ are presented here because the signals for $n\geqslant4$ are too weak and noisy. As we have mentioned, our UDV measurements can not distinguish positive (retrograde) or negative (prograde) frequencies, so the frequencies obtained from BPE are reported as absolute values. We identify the obtained components as forced flow when they are of frequency $|\omega|=1.0$, and as free modes when $|\omega| \neq 1.0$. We can see that the forced flow at $n=1$ and $n=3$ both have a constant amplitude over the entire time window, i.e. their growth rates $\alpha \approx 0$. Note that the forced flow is a superposition of the forced inertial modes which oscillate at the forcing frequency $\omega_0=1.0$ instead of their natural frequencies. We have shown in Section \ref{sec:stable} that the forced flow is dominated by the lowest inertial mode (1,1,1). In addition, we found two free modes of $n_1=1$ at $|\omega_1|=0.3410$ and $n_2=2$ at $|\omega_2|=0.6596$ with almost identical growth rate $\alpha\approx 8.8\times 10^{-3}$. Obviously, their frequencies and axial wavenumbers satisfy the triadic resonance conditions. Assuming a triadic resonance mechanism involving the forced mode (1,1,1), we identify the free modes by comparison of the observed frequency and axial wavenumber with the analytical prediction of inertial modes satisfying $m_2-m_1=1, n_2-n_1=1, \omega_2-\omega_1\approx 1.0$. The possible free modes and their inviscid frequencies $\omega_{mnk}$ are reported in the last column of Table \ref{tab:growth1_R50P03}. In this case, two inertial modes (2,1,4) and (3,2,4) are found to match the observation and the triadic resonant conditions. As we mentioned in Section \ref{inertialmode}, we note that the two free modes have the same radial structure ($k_1=k_2$),  which is optimal but not necessary. Positive $\omega_{mnk}$ indicates a retrograde mode and negative $\omega_{mnk}$ indicates a prograde mode. At $n=3$, there is another mode with frequency $|\omega|=0.3305$ and very small amplitude $B=0.04$, which may be one of the free modes of a secondary triadic resonance interacting with a higher wavenumber mode.
\begin{figure}
 \includegraphics[width=12cm]{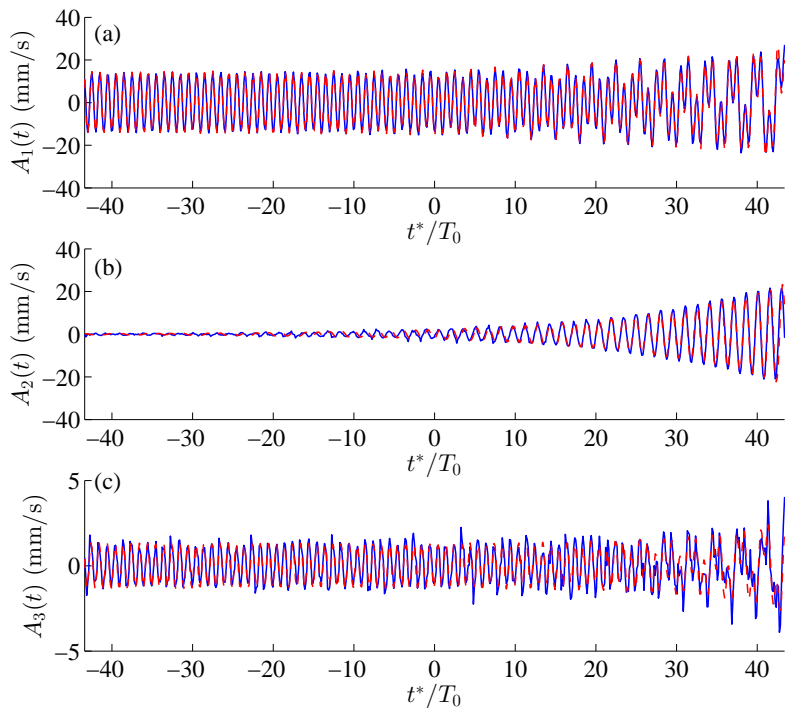} 
  \caption{Amplitudes $A_n(t)$ during the first growth phase G1 at $P_o=1.0\times10^{-2}$, $E=1.0\times10^{-5}$. The blue solid lines represent the data and the red dashed lines represent the model obtained from the BPE. Local time is used with origin in the center of the plot.} 
\label{fig:Figure7}
\end{figure}

Fig. \ref{fig:Figure7} compares the reconstructed signal with the measurements for the first growth phase at $P_o=1.0\times10^{-2}$ and $E=1.0\times10^{-5}$. We observe excellent quantitative agreement between the data and model. Although we do not have access to the azimuthal wavenumbers of the different modes to confirm $m_2- m_1=1$ because of the UDV measurement set up, the well identified axial wavenumbers and frequencies and very good agreement between the reconstructed and  
observed signals { strongly suggest} that the observed dynamics can be explained by a triadic resonance mechanism. { Meanwhile, we can not unequivocally rule out other mechanisms such as nonliner interactions in the boundary layer due to our limited observations.} Here we only show the detailed analysis of the first growth phase, the same analyses of other growth phases show a similar behaviour and the same modes interacting as in this first growth phase.
\begin{figure}
  \includegraphics[width=12cm]{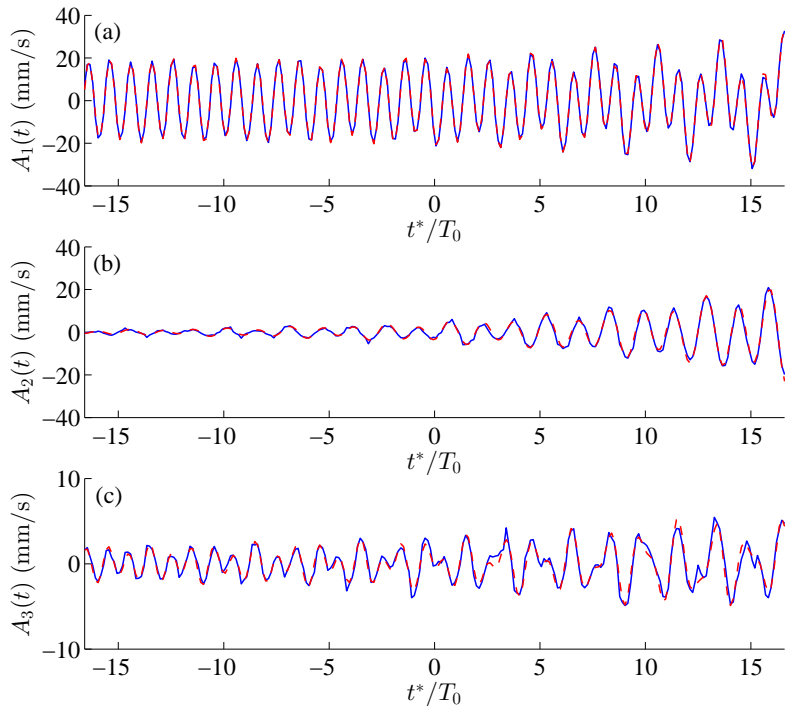} 
  \caption{Amplitudes $A_n$ during the first growth phase G1 at $P_o=1.4\times10^{-2}$, $E=1.0\times10^{-5}$. The blue solid lines represent the data and red dashed lines represent the model obtained from the BPE. Local time is used with origin in the center of the plot. }
\label{fig:Figure8}
\end{figure}
\begin{figure}
\includegraphics[width=12cm]{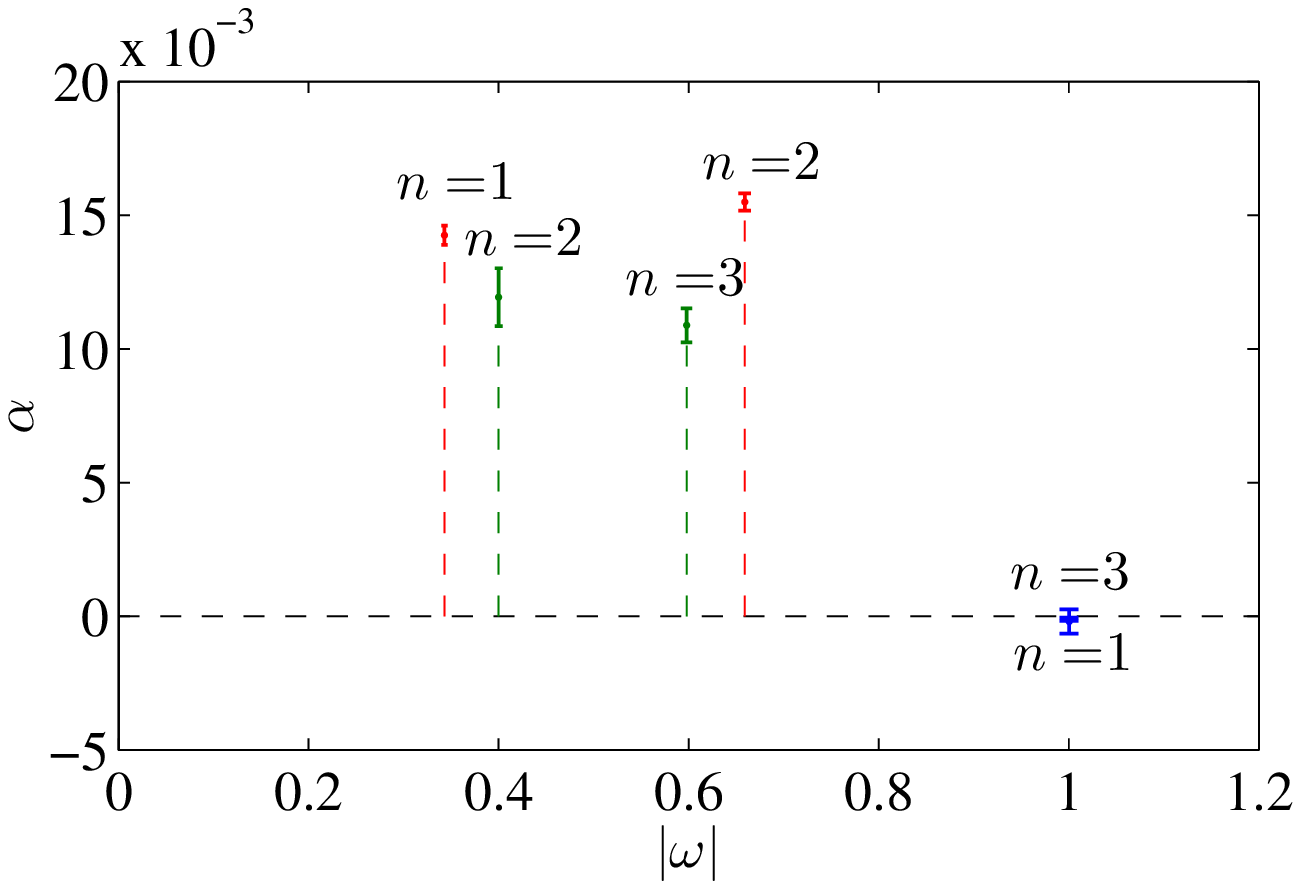}
\caption{Growth rate $\alpha$ as a function of frequency $|\omega|$ obtained from the BPE for the first growth phase G1 at $P_o=1.4\times10^{-2}$,  $E=1.0\times10^{-5}$. The frequencies and growth rates are normalized by the rotation frequency $\Omega_o$. The length of the error bars indicates uncertainties of the growth rates $\alpha$. The blue lines represent the forced flow, the red lines represent the first pair of free modes and the green lines represent the second pair of free modes.}
\label{fig:Figure9}
\end{figure}
\begin{table}
\caption{Results of the BPE for the first growth phase G1 at $P_o=1.4\times10^{-2}$,  $E=1.0\times10^{-5}$, and possible inviscid free modes matching the experimental observations.}
  \label{tab:growth1_R50P04}
  \begin{ruledtabular}
  \begin{tabular}{lrrrc}
      $n$  & Frequency ($|\omega|$) &  Growth rate ($\alpha$)  & Amplitude ($B$ (mm/s)) &   \\  [3pt]        
\hline
       1 & $1.0003(\pm6.9\times 10^{-4})$ & $-1.1\times 10^{-4}(\pm 6.8\times 10^{-5})$ & $18.73(\pm 8.7 \times 10^{-2})$ & \ \ \ Forced flow \\
       1 & $0.3430(\pm3.7\times 10^{-4})$ & $1.4\times 10^{-2}( \pm 3.6\times 10^{-4} )$ & $3.51 (\pm 7.8 \times 10^{-2})$ & \ \ Free mode $\omega_{214}=-0.3363$\\
       2 & $0.6589(\pm3.4\times 10^{-4})$ & $1.5\times 10^{-2}(\pm 3.2\times 10^{-4}) $ & $3.93 (\pm 5.4 \times 10^{-2} )$& \ \  Free mode $\omega_{324}=0.6862$\\
       2 & $0.4001(\pm1.0\times 10^{-3})$ & $1.2\times10^{-2}(\pm 1.1\times 10^{-3} )$ & $1.30( \pm 5.6 \times 10^{-2}) $ & \ \  Free mode $\omega_{327}=-0.3915$\\
       3 & $0.5978(\pm6.7\times 10^{-4})$ & $1.1 \times10^{-2}(\pm 6.4\times 10^{-4}) $ & $1.65(\pm 6.0 \times 10^{-2}) $ & \ \  Free mode $\omega_{437}=0.5963$\\
       3 & $1.0003(\pm4.8\times 10^{-4})$ & $-2.0 \times 10^{-4}(\pm 4.6\times 10^{-4}) $ & $1.69(\pm 4.8 \times 10^{-2} )$ & Forced flow\\
  \end{tabular}
  \end{ruledtabular}
\end{table}

\begin{figure}
  \includegraphics[width=12cm]{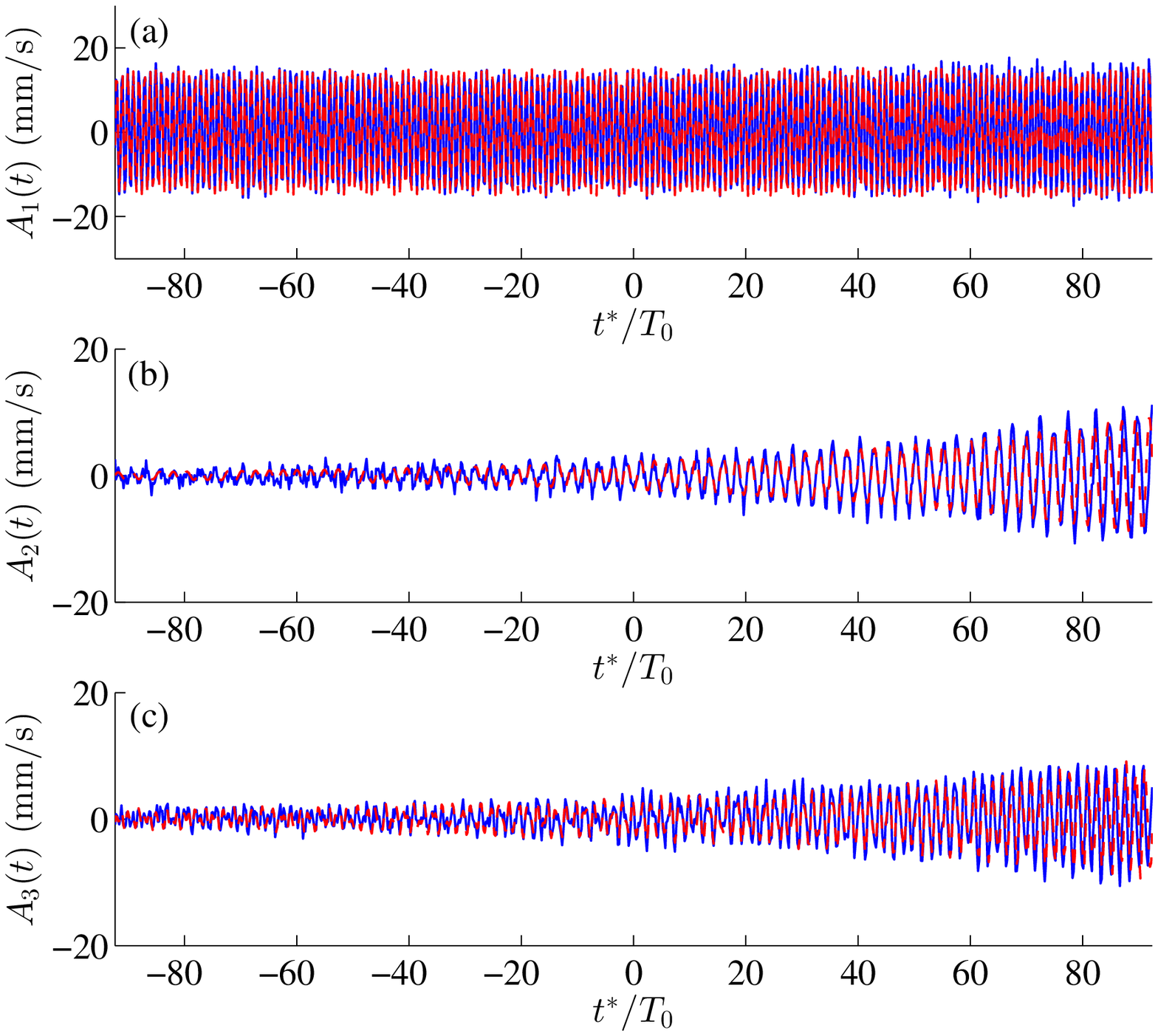}
  \caption{Amplitudes $A_n$ during the first growth phase G1 at $P_o=5.3\times10^{-3}$, $E=5.0\times10^{-6}$. The blue solid lines represent the data and the red dashed lines represent the model obtained from the BPE. Local time is used with origin in the center of the plot.  }
\label{fig:Figure10}
\end{figure}

\begin{figure}
\includegraphics[width=12cm]{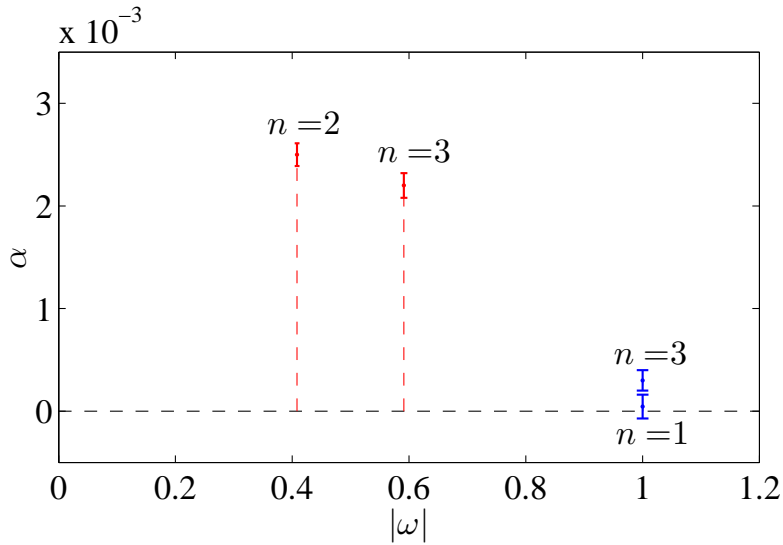}
\caption{Growth rate $\alpha$ as a function of frequency $|\omega|$ obtained from the BPE for the first growth phase G1 at $P_o=5.3\times10^{-2}$,  $E=5.0\times10^{-6}$. Only the modes with amplitude $B>1.0$ are plotted. The frequencies and growth rates are normalized by the rotation frequency $\Omega_o$. The length of the error bars indicates uncertainties of the growth rates $\alpha$. The blue lines represent the forced flow and the red lines represent the free modes.}
\label{fig:Figure11}
\end{figure}
\begin{table}
\caption{Results of the BPE for the first growth phase G1 at $P_o=5.3\times10^{-3}$,  $E=5.0\times10^{-6}$, and possible inviscid free modes matching the experimental observations. Only the modes with amplitude $B>1.0$ are listed.}
  \label{tab:growth1_R100P03}  
  \begin{ruledtabular}
  \begin{tabular}{lrrrc}
      $n$ & Frequency ($|\omega|$) & Growth rate ($\alpha$) &  Amplitude ($B$ (mm/s)) &\\  [3pt]    
\hline
       1 & $1.0001(\pm1.1\times 10^{-4})$ & $4.5\times 10^{-5}(\pm 1.2\times 10^{-4})$ & $15.02(\pm7.8\times 10^{-2})$ & Forced flow\\
       2 & $0.4083(\pm1.1\times 10^{-4})$ & $2.5\times 10^{-3}(\pm 1.1\times 10^{-4})$ & $2.25(\pm3.7\times 10^{-2})$ & Free mode $\omega_{327}=-0.3915$\\
       3 & $0.5913(\pm2.7\times 10^{-4})$ & $2.2 \times 10^{-3}(\pm 1.2\times 10^{-4})$ & $2.36( \pm 2.9\times 10^{-2})$ & Free mode $\omega_{437}=0.5963$\\
       3 & $1.0000(\pm1.0\times 10^{-4})$ & $3.0\times 10^{-4}(\pm 1.0\times 10^{-4})$ & $1.20(\pm1.9\times 10^{-2})$ & Forced flow\\
  \end{tabular}
   \end{ruledtabular}
\end{table}

In some cases, we observe more than one single triadic resonance during a single growth phase. This is illustrated by the results at $P_o=1.4\times10^{-2}$ and $E=1.0\times10^{-5}$. The time series  $A_n(t)$ during the first growth phase are represented in Fig. \ref{fig:Figure8}. The modes obtained using the BPE for $n\leqslant 3$ are plotted in Fig. \ref{fig:Figure9} and listed in Table \ref{tab:growth1_R50P04}. As previously, we observe a triadic resonance between the forced flow and the two free modes of $n_1=1$ at $|\omega_1|=0.3430$ and $n_2=2$ at $|\omega_2|=0.6589$ with a similar growth rate $\alpha \approx1.4\times 10^{-2}$. The same free modes have been observed in the previous case at $P_o=1.0\times10^{-2}$ and $E=1.0\times10^{-5}$. In addition, we observe another pair of free modes of $n_3=2$ at $|\omega_3|=0.4001$ and $n_4=3$ at $|\omega_4|=0.5978$ with a similar growth rate $\alpha\approx 1.2\times 10^{-2}$, which also satisfy the triadic resonance conditions with the forced mode (1,1,1). The corresponding inviscid modes should be (3,2,7) and (4,3,7) as shown in Table \ref{tab:growth1_R50P04}. We note that the growth rate of the first triadic resonance is larger than the second one. It is the first time that two triadic resonances occurring simultaneously have been observed experimentally. Again, the growth rate of the forced flow is almost zero. As previously the reconstructed signals with the modes identified using the BPE are in excellent quantitative agreement with the observations (Fig. \ref{fig:Figure8}).

Considering now lower Ekman numbers, Fig. \ref{fig:Figure10} shows an example of a triadic resonance during the first growth phase at Ekman number $E=5.0\times 10^{-6}$ and $P_o=5.3\times10^{-3}$. The corresponding results of the BPE are plotted in Fig. \ref{fig:Figure11} and listed in Table \ref{tab:growth1_R100P03}. We can see that there is no free mode with $n=1$ in this case. Only one dominant triadic resonance is observed involving the free modes $n_1=2$ at $|\omega_1|=0.4080$ and $n_2=3$ at $|\omega_2|=0.5913$. The two free inertial modes are identified as (3,2,7) and (4,3,7) (see Table \ref{tab:growth1_R100P03}), which correspond to the same free inertial modes of the second triadic resonance at $P_o=1.4\times10^{-2}$ and $E=1.0\times10^{-5}$. Again, the reconstructed signals with modes listed in Table \ref{tab:growth1_R100P03} are in excellent agreement with the observations. 

By comparing the previous three cases of triadic instability, we notice that the most unstable free modes depend on the Ekman number. At the Ekman number $E=1.0\times10^{-5}$, the most unstable free modes are (2,1,4) and (3,2,4), while at lower Ekman number $E=5.0\times10^{-6}$, the most unstable free modes are (3,2,7) and (4,3,7), which have higher wavenumbers. This is in agreement with the global stability theory of inertial modes \cite{Kerswell1999, Lagrange2011}, which can be understood as follows. For asymptotically small Ekman numbers, the viscous damping and the peak of the eigen-frequency of the inertial modes scale as $E^{1/2}$. Hence, there is a competition between the viscous damping that tends to reduce the growth rate of large wave number modes at moderate Ekman numbers while the frequency detuning effect allows more pairs of modes to be coupled  \cite{Kerswell1999}. At low Ekman numbers, it is to the contrary, the viscous damping is reduced but the detuning effect is smaller preventing certain pairs of free modes to be in resonance with the forced flow.  Consequently, as the Ekman number is lowered some triadic resonance can become secondary and eventually sub-critical when the viscous detuning is no longer sufficient to couple the modes.

We also notice that the observed frequencies of two free modes are all smaller than 1.0, i.e. $|\omega_1|+|\omega_2|=1.0$. We never observe two free modes satisfy $|\omega_2|-|\omega_1|=1.0$ or $|\omega_1|-|\omega_2|=1.0$, i.e. the absolute value of frequency of one free mode is larger than 1.0 and the other is smaller than 1.0. This can be explained by the dispersion relation of inertial modes (See Appendix \ref{app:fre}).

\subsubsection{Collapse phase}

\begin{figure}
  \includegraphics[width=15cm]{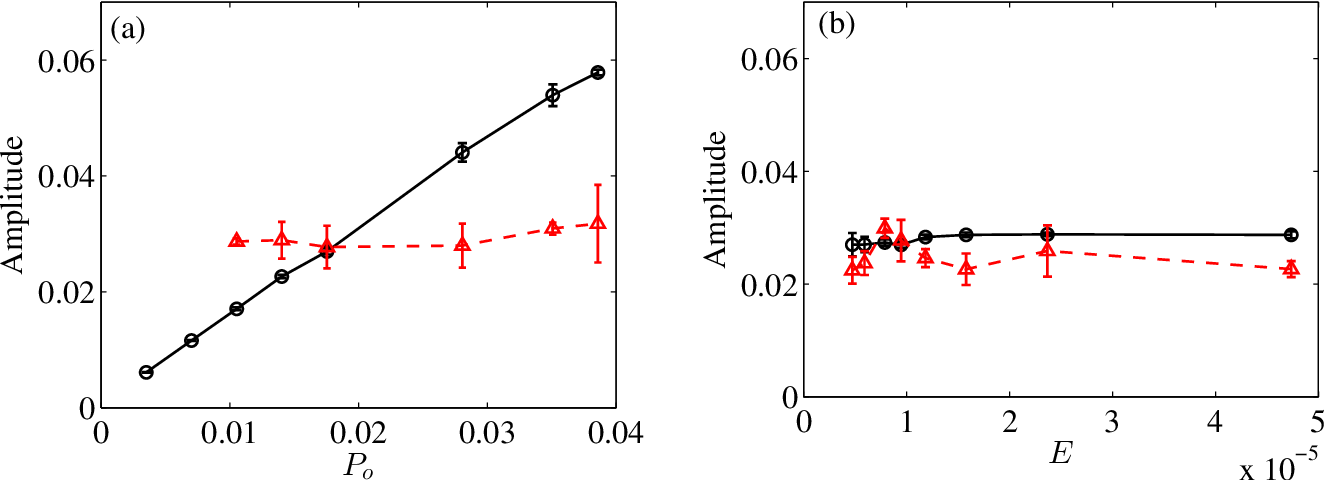}
  \caption{Nondimensional amplitude of the forced flow $n=1$, $\omega=1.0$ (black) and the free mode $n=2$, $\omega=0.66$ (red) just before the collapse (a) as a function of  $P_o$ at fixed $E=1.0\times10^{-5}$ and (b) as a function of $E$ at fixed $P_0=1.75\times10^{-2}$. The amplitude range is set to be the same in (a) and (b). The length of the error bars indicates the standard deviation of amplitudes obtained from different growth-collapse events.}
\label{fig:Figure12}	
\end{figure}

\begin{figure}
\includegraphics[width=15cm]{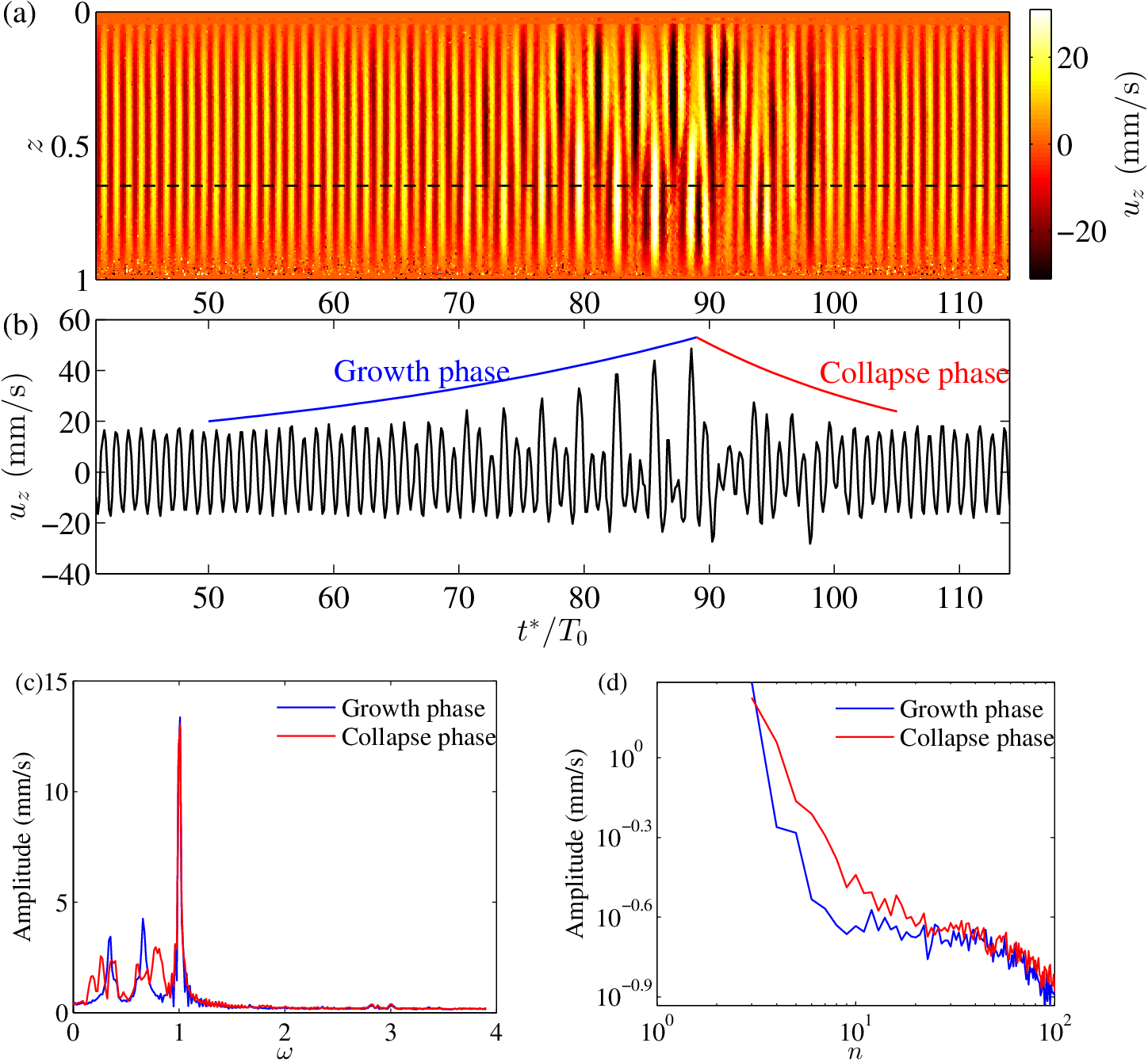}
\caption{The first growth-collapse event at $P_o=1.4\times10^{-2}$ and $E=1.0\times10^{-5}$. (a) UDV measurements of axial velocity as a function of time and $z$. (b) Time series of the axial velocity at $z=0.65$ (black dashed line in (a)). The blue line and red line schematically show the growth phase and collapse phase respectively. (c) Amplitude of the DFT in time averaged over $z$ during the growth phase (blue line) and collapse phase (red line).
(d) Amplitude of DFT as a function of axial wavenumber $n$ averaged over time during the growth phase (blue line) and collapse phase (red line).
}
\label{fig:Figure13}
\end{figure}

In all of our experiments, the free modes grow until their amplitudes reach a critical value and collapse to a more complex flow. This so-called resonant collapse was reported in previous experimental studies\cite{McEwan1970,Manasseh1992,Kobine1995,Kobine1996}. 
Several important questions remain open regarding the underlying mechanism of the collapse. 

Fig \ref{fig:Figure12} shows the nondimensional amplitude of the forced flow and of the free mode of $n=2$ and $|\omega|=0.66$ just before the collapse as a function of the Poincar\'e number $P_o$ and the Ekman number $E$. 
As expected from asymptotic theory\cite{Liao2012}, far from a primary resonance, the amplitude of the forced flow is proportional to $P_o$ and independent of $E$. The  amplitude of the free modes grows until it reaches a critical value of about $0.03$, independent of $Po$ and $E$. Above this critical value the free modes collapse. In addition, we note that, at low Poincar\'e numbers, the maximum amplitude of the free mode can be comparable and even larger than the amplitude of the forced flow.
 
Fig. \ref{fig:Figure13} shows an example of growth-collapse event at  $P_o=1.4\times10^{-2}$ and $E=1.0\times10^{-5}$. The UDV measurements of the axial velocity during the first growth-collapse event are plotted as a function of time and depth in Fig. \ref{fig:Figure13} (a). Fig.  \ref{fig:Figure13} (b) shows the time series of the axial velocity at $z=0.65$ corresponding the black dashed line in Fig. \ref{fig:Figure13} (a). The growth phase has been well explained by the triadic resonance mechanism. The exponential growth suddenly collapses leading to a more complex flow when the amplitude of the free modes reaches a critical value. 
In Fig. \ref{fig:Figure13} (c) we compute the DFT in time averaged over depth $z$ during the growth phase ($50\leqslant t^*/T_0\leqslant87$, blue curve) and during the collapse phase ($88\leqslant t^*/T_0\leqslant108$, red curve). During the collapse phase, the amplitude of the forced flow ($\omega=1.0$) does not change significantly compared to the growth phase. The frequency spectrum during the collapse phase exhibits numerous peaks at frequencies close to those of the free modes during the growth phase. In contrast with the case of large precession rate (Fig. \ref{fig:Figure14} (c)), the flow remains in the frequency range of [0, 1.2], which is within the inertial wave frequency range. This may suggest that the flow is still in the form of inertial waves or modes during the collapse phase. Fig. \ref{fig:Figure13} (d) shows the spatial DFT in $z$ averaged over the growth time window (blue curve) and collapse time window (red curve). The large scale forced flow is still maintained during the collapse phase, so the axial wavenumber spectrum only for $n\geqslant 3$ is plotted here to see more details about small scales. We observe a slight increase of the DFT amplitude in the range of $n\in[3,20]$ during the collapse phase. We note from Fig \ref{fig:Figure13} (b) that the collapse time duration is of order $10T_0$. The free decay time scale of the inertial mode, i.e. the inverse of the decay rate, can be calculated analytically in a cylinder considering both viscous decay in the boundary layer and in the interior \cite{Zhang2008a}. There is no similar expression available in a cylindrical annulus with moderate radius ratio. Here we numerically calculate the decay rate of the inertial mode in a cylindrical annulus using the method in Ref \onlinecite{Kerswell1995a}. The decay rate and decay time scale of the inertial modes with typical wavenumber of 1, 10 and 20 are listed in Table \ref{tab:decayrate}. We can see that the collapse time duration is comparable to the free decay time of the mode (10,10,10) and (20,20,20). This suggests that during the collapse phase the energy may be transferred to smaller scale inertial modes that are more efficient at dissipating energy.

\begin{table}
\caption{\label{tab:decayrate} Decay rates of some inertial modes calculated using the numerical method in Ref \onlinecite{Kerswell1995a}.}
\begin{ruledtabular}
\begin{tabular}{lcccc}
 &\multicolumn{2}{c}{$E=1.0\times10^{-5}$}&\multicolumn{2}{c}{$E=5.0\times 10^{-6}$}\\
 ($m,n,k$)&Decay rate&Decay time scale &Decay rate & Decay time scale \\ \hline
 (1,1,1)&$1.10\times 10^{-2}$& $91 T_0$ &$7.72\times10 ^{-3}$&$130 T_0$ \\
 (10,10,10)&$4.08\times 10^{-2}$& $25 T_0$ &$2.23\times10 ^{-3}$&$45 T_0$ \\
 (20,20,20)&$1.37\times 10^{-1}$& $7 T_0$ &$7.02\times10 ^{-2}$&$14 T_0$ \\
\end{tabular}
\end{ruledtabular}
\end{table}

\subsection{Large precession rate } \label{sec:large}
\begin{figure}
  \includegraphics[width=15cm]{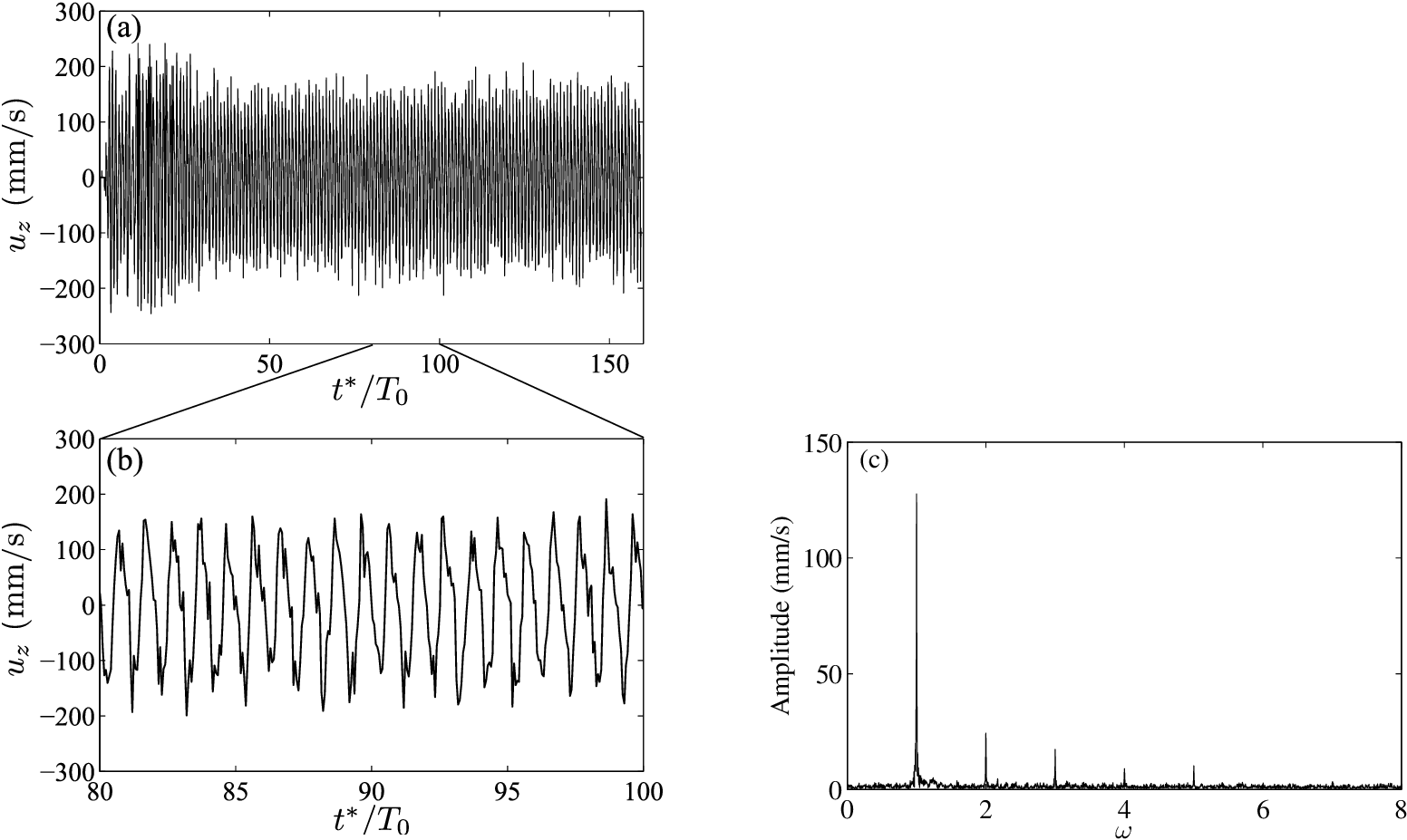}
  \caption{(a) Time series of the axial velocity at $z=0.35$ recorded by the UDV. (b) Zoom in of (a) between $t^*/T_0=80$ and $t^*/T_0$=130. (c) Amplitude of the DFT of the time series (a). $P_o=0.1$, $E=1.0\times10^{-5}$.}
\label{fig:Figure14}
\end{figure}
At large enough Poincar\'e number we observe different dynamics without any evidence of triadic resonances as illustrated in Fig. \ref{fig:Figure14}. The velocity time series (Fig.  \ref{fig:Figure14} (a) and (b)) and frequency spectrum (Fig. \ref{fig:Figure14} (c)) are characterized by a forced oscillation at $\omega=1$ and multiple harmonics of smaller amplitude with frequency $\omega=2.0$, $\omega=3.0$, $\omega=4.0$ and so on. Such a flow can no-longer be in the form of inertial modes which exist only in the range $|\omega|\leqslant 2$. This type of spectrum is typical of strong non-linear interactions which are not strongly influenced by the rotation. 
In addition above $\omega=4.0$, the UDV reaches its limits as the typical sampling period becomes comparable to the typical fluctuation time scales of the velocity. 
In the context of rapidly rotating object such as planetary cores, this range of parameters  is of limited interest and we did not intend to investigate it in details.

\subsection{Instability diagram}\label{sec:diagram}
\begin{figure}
\includegraphics[width=10cm]{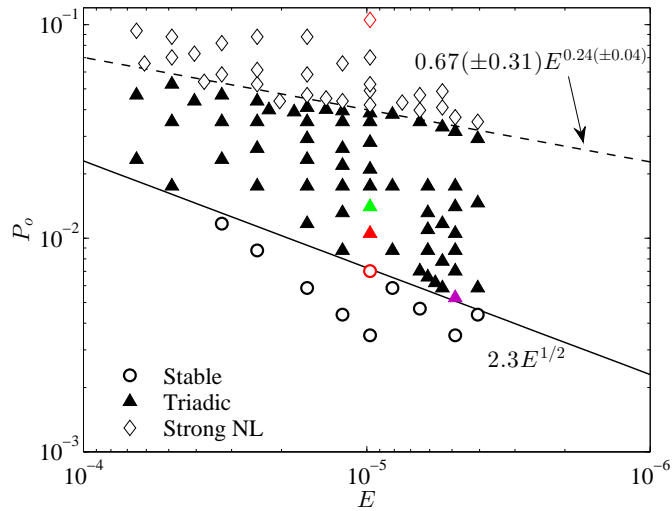}
\caption{Instability diagram of the fluid flow in this experimental study. Open circles denote the stable flow described in Section \ref{sec:stable}. Filled triangles denote the flow characterized as the triadic resonance (Section \ref{sec:triadic}). Open diamonds denote the the large $P_o$ regime described in in Section \ref{sec:large}. The solid line represents the predicted threshold of the triadic resonance instability ($P_{oc}\sim E^{1/2}$). {The dashed line represents the least square fitting of power law using points around the second transition.} The colored symbols denote the corresponding experiments discussed in detail. The red circle corresponds to Fig. \ref{fig:Figure3}, the red triangle corresponds to Figs. \ref{fig:Figure4}-\ref{fig:Figure7}, the green triangle corresponds to Figs. \ref{fig:Figure8} and \ref{fig:Figure9}, the violet triangle corresponds to Figs. \ref{fig:Figure10} and \ref{fig:Figure11}, and the red diamond corresponds to Fig. \ref{fig:Figure14}.}
\label{fig:Figure15}
\end{figure}

We have characterized three different flow regimes by varying the Poincar\'e number $P_o$ and Ekman number $E$. Fig. \ref{fig:Figure15} shows the instability diagram of the fluid flow in the ($E,\, P_o$) plane. For each combination of ($E,\, P_o$), the flow is characterized as stable (Section \ref{sec:stable}) or triadic resonance (Section \ref{sec:triadic})  or strong nonlinear regime (Section \ref{sec:large}) based on the UDV measurements. 
As we have mentioned, the inviscid growth rate of the triadic resonance is proportional to the amplitude of the forced flow which is $O(P_o)$ in the tank far from a primary resonance (Fig. \ref{fig:Figure12}). The inviscid growth rate is reduced by the viscous damping and detuning which is $O(E^{1/2})$. Therefore, the critical Poincar\'e number of the onset of the triadic resonance  is given as $P_{oc}=O(E^{1/2})$ (the solid line in Fig. \ref{fig:Figure15}). We can see that this scaling is in good agreement with the experimental results. At large $P_o$, we do not know the underlying mechanism of this flow regime, and the theoretical prediction of the second threshold is not available. {So we do the least square fitting of a power law $P_{oc}=\beta E^{\gamma}$ using points around the second transition, which gives $P_{oc}=0.67(\pm0.31)E^{0.24(\pm0.04)}$ (the dashed line in Fig \ref{fig:Figure15}) with uncertainties in the bracket.} 
          
\section{Discussion}\label{Diss}
In the present study, we have shown that the dynamics driven by precession in a cylindrical annulus of moderate radius ratio is similar to that of a full cylinder. At weak precession, the fluid flow is stable and well described by a superimposition of inertial modes. 

At a given Ekman number, there exists a critical Poncar\'e number above which the forced flow can excite free modes {, the axial wavenumbers and frequencies of which satisfy the triadic resonance conditions.}
Using a Bayesian analysis, we extracted the frequencies, growth rates, amplitudes of modes with different wavenumbers during the growth phases.
In some cases, we observed that two pairs of free modes both satisfying the triadic resonance conditions can grow simultaneously. This has been predicted by previous theoretical and numerical studies\cite{Kerswell1999, Mason1999}, but, to our knowledge, it is the first experimental evidence. 
Our experiments have shown that the amplitude of the forced flow does not vary significantly, i.e. almost zero growth rate, while the free modes are exponentially growing. This suggests that the energy it exchanges with the free modes through the triadic resonance mechanism is instantaneously recovered. Hence, the forced flow acts as a ``carrier", the energy being provided ultimately by the motor and stored in the free modes during the growth phase and dissipated during the collapse phase.
     
The exponential growth of the free modes invariably collapse when their amplitudes reach a critical value which is about 3\% of $\Omega_o R_o$(see Fig. \ref{fig:Figure12}). The phenomenon is known as resonant collapse and is named after McEwan\cite{McEwan1970}. 
Different scenarios have been discussed in the context of the elliptical instability in Ref \onlinecite{Eloy2003}. These authors excluded the boundary layer instability or centrifugal instability and argued that the resonant collapse is likely due to the nonlinear interaction of several inertial modes. It is still impossible for us to draw a hard conclusion about the mechanism of the collapse, but we have learnt a few facts based on our experiments. Firstly, the maximum amplitude of the free mode just before the collapse, i.e. the critical amplitude,  is independent of the Poincar\'e number and Ekman number (Fig. \ref{fig:Figure12}). Secondly, the maximum amplitude of the free mode can be comparable and even a little bit larger than the amplitude of the forced flow. These two points suggest that the free modes themselves are inertially unstable and the forced flow plays a passive role in the collapse. Thirdly, the frequency spectrum during the collapse phase shows that the energy remains in the frequency range of inertial wave. Furthermore, the spatial DFT shows a small increase of the energy in the axial wavenumber range of [3,20] and the collapse time duration is comparable with the free decay time of inertial modes with wavenumbers of 10 and 20.
These facts suggest that the flow during the collapse phase is still in the form of inertial waves or modes which could be the product of a subsequent triadic resonance, i.e. the free mode plays the role of the forced mode and excite two other free modes as proposed by Ref \onlinecite{Mason1999}. 
The energy stored in the free modes during the growth phase is dissipated by the smaller scale inertial modes during the collapse phase.

\begin{acknowledgments}
This work grew from stimulating discussions with Keke Zhang, whose input is gratefully acknowledged. We also thank Jose Abreu for help in carrying out  the Bayesian analysis and Peter Scarfe, Roland Grimmer and Thomas M\"orgeli for technical support on the experiment. This study is financially supported by ERC Grant No. 247303 (MFECE).
\end{acknowledgments}

\appendix
\section{Inertial modes in a cylindrical annulus}\label{App1}
Eliminating the velocity $\bm u$ in the Eqs. (\ref{eq:nseq_1}) and (\ref{eq:incompressible_1}), we get the Poincar\'e equation in terms of the pressure $p$ describing inertial waves or inertial modes\cite{Greenspan1968}. In cylindrical coordinates $(r,\phi,z)$ with the $z$ axis as the rotation axis, the Poincar\'{e} equation can be written as \cite{Greenspan1968}
\begin{equation}\label{Aeq1}
\frac{1}{r}\frac{\partial p}{\partial r} +\frac{1}{r^2}\frac{\partial^2 p}{\partial r^2}+\frac{1}{r^2}\frac{\partial^2 p}{\partial \phi^2}-\Big(\frac{4-\omega^2}{\omega^2}\Big)\frac{\partial ^2p}{\partial z^2}=0, 
\end{equation}
with boundary conditions on the solid walls\cite{Greenspan1968}
\begin{equation}\label{Aeq2}
4(\boldsymbol{\hat z}\cdot \nabla p)(\boldsymbol{\hat z}\cdot \boldsymbol{\hat n})-\omega^2\boldsymbol{\hat n}\cdot \nabla p-2\mathrm{i} \omega (\boldsymbol{\hat n}\times \boldsymbol{\hat z})\cdot \nabla p=0,
\end{equation}
where $\omega$ is the frequency of inertial modes bounded $|\omega|\leqslant 2$ and $\boldsymbol{\hat n}$ indicates the outer unit vector of the boundary surface. The velocity field can be obtained from the pressure $p$ as\cite{Greenspan1968}
\begin{equation}\label{Aeq3}
\boldsymbol{u}=\frac{\mathrm{i}}{\omega(4-\omega^2)}[4(\boldsymbol{\hat z}\cdot \nabla p)\boldsymbol{\hat z}-\omega^2\nabla p- 2\mathrm{i}\omega \boldsymbol{\hat z}\times \nabla p].
\end{equation}

In a cylindrical annulus with aspect ratio $h$ and radius ratio $r_{i}$, the Poincar\'e equation can be solved using separation of variables. The general solution of the equation is
\begin{equation}\label{Aeq4}
p(r,z,\phi)=[C^1_{mn}J_m({\xi r})+C^2_{mn}Y_m(\xi r)]\cos\Big(\frac{n\pi z}{h}\Big)\mathrm e^{\mathrm{i}m\phi}, \quad m=0,1,2,..., n=1,2,3,...,
\end{equation}
where $J_m$ and $Y_m$ are the first and second kind of Bessel Function for integer order $m$. $C^1_{mn}$ and $C^2_{mn}$ are constants for given $m$, $n$ and $\xi$. The integer $m$ and $n$ is azimuthal wavenumber and axial wavenumber respectively. $\xi$ is radial wavenumber and determined by the boundary conditions.

\begin{figure}
\includegraphics[width=15cm]{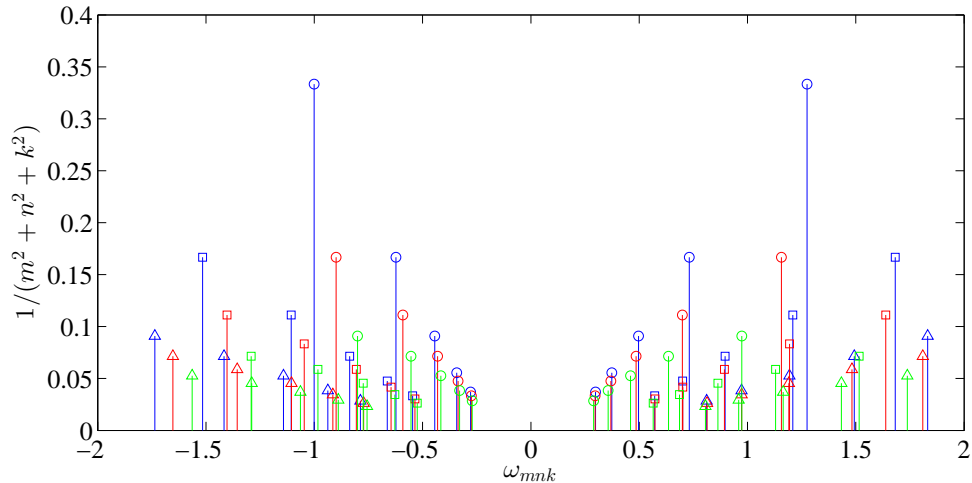}
\caption{Frequency spectrum of inertial modes with $m=1,\,2,\,3$, $n=1,\,2,\,3$ and $k\leqslant 5$ in a cylindrical annulus with the aspect ratio $h=1.0$ and radius ratio $r_i=0.269$. The blue lines represent the inertial modes with $m=1$, the red lines represent the inertial modes with $m=2$ and the black lines represent the inertial modes with $m=3$. The circles represent the inertial modes with $n=1$, the squares represent the inertial modes with $n=2$ and the triangles represent the inertial modes with $n=3$. The vertical axis is a measure of the quality factor (inverse damping) of the mode. The retrograde mode $\omega_{134}=0.9722$ close to $\omega=1.0$ is not resonant with precession provided $E<10^{-4}$. The prograde modes (negative frequencies) are never resonant with precession because the precessional force is retrograde in the rotating frame.}
\label{fig:Figure16}
\end{figure}

Imposing the boundary conditions  at $r=r_{i}$ and $r=1$, we get 
\begin{equation}\label{Aeq5}
[\omega \xi J_m'(\xi)+2mJ_m(\xi)]C^1_{mn}+[\omega \xi Y_m'(\xi)+2mY_m(\xi)]C^2_{mn}=0,
\end{equation}
\begin{equation}\label{Aeq6}
[r_{i}\omega \xi J_m'(r_{i}\xi)+2mJ_m(r_{i}\xi)]C^1_{mn}+[r_{i}\omega \xi Y_m'(r_{i}\xi)+2mY_m(r_{i}\xi)]C^2_{mn}=0.
\end{equation}
In order to get no-trivial solutions, $C^1_{mn}$ and $C^2_{mn}$ should not always equal zero. So the determinant of the coefficient matrix should be equal to zero
\begin{equation}\label{det}
\left|\begin{array}{cc}
\omega \xi J_m'(\xi)+2mJ_m(\xi) & \omega \xi Y_m'(\xi)+2mY_m(\xi)\\
r_{i}\omega \xi J_m'(r_{i}\xi)+2mJ_m(r_{i}\xi) & r_{i}\omega \xi Y_m'(r_{i}\xi)+2mY_m(r_{i}\xi)
\end{array}\right|=0.
\end{equation}
For simplicity, we define
\begin{eqnarray}\label{eq:dingyi}
P_m=J_m(\xi)Y_m({r_{i}\xi})-J_m(r_{i}\xi)Y_m({\xi}), \\
Q_m=J_m(\xi)Y_m'({r_{i}\xi})-J_m'(r_{i}\xi)Y_m({\xi}), \\
R_m=J_m'(\xi)Y_m({r_{i}\xi})-J_m(r_{i}\xi)Y_m'({\xi}), \\
S_m=J_m'(\xi)Y_m'({r_{i}\xi})-J_m'(r_{i}\xi)Y_m'({\xi}),
\end{eqnarray}
which have the following recurrence relations\cite{Abramowitz1964}
\begin{eqnarray}
P_{m+1}-P_{m-1} = -\frac{2m}{\xi}Q_m-\frac{2m}{r_{i}\xi}R_m, \\
S_m =\frac{1}{2}P_{m+1}+\frac{1}{2}P_{m-1}-\frac{m^2}{r_{i}\xi^2}P_m.
\label{eq:rec}
\end{eqnarray}
Plugging Eq. (\ref{eq:dingyi}-\ref{eq:rec}) into the Eq. (\ref{det}), we get the transcendental equation 
\begin{equation}\label{traneq}
r_{i}\xi ^2 (P_{m+1}+P_{m-1})-2 r_{i}\frac{\xi^2}{\omega}(P_{m+1}-P_{m-1})+2\Big(\frac{\xi h}{n \pi}\Big)^2m^2P_m=0.
\end{equation}
There are an infinite number of solutions of the equation for given $m$ and $n$, and the positive solutions are arranged in ascending order 
\begin{equation}
\xi_{mn1}<\xi_{mn2}<...<\xi_{mnk}<... .
\end{equation}
where $\xi_{mnk}$ is the $k$th positive solution. Generally, the  radial velocity of the inertial mode corresponding to $\xi_{mnk}$ has $k$ nodes in radial direction, so we refer $k$ as to the integer radial wavenumber. The eigen-frequency of the inertial mode with wavenumbers ($m,n,k$) is then given by the dispersion relation\cite{Greenspan1968} 
\begin{equation}\label{eq:dispersion}
\omega_{mnk}^2=\frac{4n^2\pi^2}{(\xi_{mnk}h)^2+(n\pi)^2}.
\end{equation}
Positive $\omega_{mnk}$ indicates a retrograde mode and negative $\omega_{mnk}$ indicates a prograde mode. Fig. \ref{fig:Figure16} shows the eigen-frequencies of low order inertial modes in a cylindrical annulus with the aspect ratio $h=1.0$ and $r_i=0.269$.  
 
Finally, the velocities of the inertial modes are obtained from Eq. (\ref{Aeq3})
\begin{eqnarray}
\boldsymbol{\hat r}\cdot \boldsymbol{u}_{mnk} &= &-\mathrm{i}\Big[\omega_{mnk}\xi_{mnk}\big(C^1_{mnk}J_{m-1}(\xi_{mnk}r)+C^2_{mnk}Y_{m-1}(\xi_{mnk}r)\big) \nonumber\\
 & & +\frac{m(2-\omega_{mnk})}{r}\big(C^1_{mnk}J_{m}(\xi_{mnk}r)+C^2_{mnk}Y_{m}(\xi_{mnk}r)\big)\Big] \\
 & & \times \Big(\frac{1}{4-\omega_{mnk}^2}\Big)\cos(n\pi z/h)\mathrm{e}^{\mathrm{i}m\phi} \nonumber,
\end{eqnarray}
\begin{eqnarray}
\boldsymbol{\hat \phi}\cdot \boldsymbol{u}_{mnk}&= &
\Big[2\xi_{mnk}\big(C^1_{mnk}J_{m-1}(\xi_{mnk}r)+C^2_{mnk}Y_{m-1}(\xi_{mnk}r)\big)  \nonumber\\
 & & -\frac{m(2-\omega_{mnk})}{r}\big(C^1_{mnk}J_{m}(\xi_{mnk}r)+C^2_{mnk}Y_{m}(\xi_{mnk}r)\big)\Big] \\
 & & \times \Big(\frac{1}{4-\omega_{mnk}^2}\Big)\cos(n\pi z/h)\mathrm{e}^{\mathrm{i}m\phi} \nonumber,
\end{eqnarray}

\begin{equation}\label{eq:AppUz}
\boldsymbol{\hat z}\cdot \boldsymbol{u}_{mnk}=\frac{-\mathrm{i}n\pi}{h\omega_{mnk}}\Big[C^1_{mnk}J_m(\xi_{mnk}r)+C^2_{mnk}Y_m(\xi_{mnk}r)\Big]
\sin(n\pi z/h)\mathrm{e}^{\mathrm{i}m\phi},
\end{equation}
where
\begin{equation}
C^1_{mnk}=C_0[\omega_{mnk}\xi_{mnk} Y_m'(\xi_{mnk})+2mY_m(\xi_{mnk})],
\end{equation}
\begin{equation}
C^2_{mnk}=-C_0[\omega_{mnk}\xi_{mnk} J_m'(\xi_{mnk})+2mJ_m(\xi_{mnk})],
\end{equation}
and $C_0$ is an arbitrary constant.  
Suitably normalized, inertial modes satisfy the following orthogonality\cite{Greenspan1968}
\begin{equation}
\int \boldsymbol{u}_{mnk}\cdot \boldsymbol{u}_{m'n'k'}^* \mathrm{d}V=\delta_{mm'}\delta_{nn'}\delta_{kk'},
\end{equation} 
where$\boldsymbol{u}_{m'n'k'}^*$ is the complex conjugate of $\boldsymbol{u}_{m'n'k'}$ and $\delta$ represents the Kronecker delta function. 
\section{Linear inviscid theory}\label{App2}
\begin{table}
\caption{Amplitude (and frequency) of each mode given by Eq. (\ref{eq:B4}). $P_o=1.0$, $r_{i}=0.269$, $h=1.0$.}
  \label{tab:InviscidAmp}
  \begin{ruledtabular}
  \begin{tabular}{lrrrrr}
    & \ \  $k=1$ \ \    & \ \ $k=2$ \ \  &\ \  $k=3$ \ \  & \ \ $k=4$ \ \  &\ \  $k=5$ \ \  \\  [3pt]      
\hline
$n=1$\ \ & 0.5306 (1.2748 )  &  -0.2072 (0.7312) &  0.1926 (0.4969) &  -0.0108 (0.3737) & -0.0476 (0.2987) \\
$n=2$\ \ & 0.0 ( 1.6822)  &  0.0 (1.2093) &  0.0 (0.8965)&  0.0 (0.7006) & 0.0 (0.5714)\\
$n=3$\ \ & -0.0030 (1.8321)  & 0.0094 (1.4932)  &  -0.0367 (1.1938)  & 0.0335 (0.9722) & -0.0292 (0.8117) \\
$n=4$\ \ & 0.0  (1.8986) &  0.0 (1.6593) &  0.0 (1.4047) &  0.0 (1.1880) & 0.0 (1.0162) \\
$n=5$\ \ & 0.0012  (1.9328) &   -0.0027 (1.7597)&   -0.0010 (1.5516) &   0.0094 (1.3546) &   -0.0033 (1.1853) \\
  \end{tabular}
  \end{ruledtabular}
\end{table}

The linear inviscid inertial modes excited by precession are governed by the following equations
\begin{eqnarray}\label{eq:inv_NS}
   \frac{\partial \boldsymbol u}{\partial t}+2\boldsymbol{\hat{z}}\times \boldsymbol u &=&  -\nabla p-2\boldsymbol{\hat{z}}P_o r\mathrm{e}^{\mathrm{i}(t+\phi)},  \\
   \nabla \cdot \boldsymbol u& =&  0,
\end{eqnarray}
where the real part of the solution represents the physical solution. The precession force $ -2\boldsymbol{\hat{z}}P_o r\mathrm{e}^{\mathrm{i}(t+\phi)}$ has azimuthal wavenumber $m=1$ and dimensionless frequency $\omega=1$, so the solution can be expanded as \cite{Liao2012}
\begin{equation}
\boldsymbol{u}=\sum_{n}\sum_{k}\mathcal{A}_{1nk}\boldsymbol{u}_{1nk}\mathrm e^{\mathrm it}, \quad p=\sum_{n}\sum_{k}\mathcal{A}_{1nk}p_{1nk}\mathrm e^{\mathrm it}.
\label{eq:B3}
\end{equation}  
Substituting Eq. (\ref{eq:B3}) into the Eq. (\ref{eq:inv_NS}) and using the orthogonality of the inertial modes, we get the amplitude of each mode \cite{Greenspan1968}
\begin{equation}
\mathcal{A}_{1nk}=\frac{\int (-2\mathbf{\hat z}P_o re^{\mathrm i \phi})\cdot \boldsymbol{u}_{1nk}^* \mathrm{d} V }{\mathrm i(1-\omega_{1nk})\int \boldsymbol{u}_{1nk}\cdot \boldsymbol{u}_{1nk}^* \mathrm{d}V},
\label{eq:B4}
\end{equation}
where $\boldsymbol{u}_{1nk}^*$ is the complex conjugate of $\boldsymbol{u}_{1nk}$. Table \ref{tab:InviscidAmp} lists the amplitudes of inertial modes with $n\leqslant 5$ and $k\leqslant 5$ in a cylindrical annulus with aspect ratio $h=1.0$ and radius ratio $r_{i}=0.269$ corresponding the one we use in the experiments. Due to the parity of the precession force in the axial direction, precession can only excite the modes with odd axial wavenumber $n$ as we can see from the table.

\section{Frequencies of the two free inertial modes}{\label{app:fre}}
In all our experiments, the absolute value of the frequencies of two free modes are all smaller than 1.0. This can be explained as follows.
The absolute value of the frequency of the inertial mode is given as (Eq. \ref{eq:dispersion})
\begin{equation}
|\omega_{mnk}|=\frac{2 n \pi}{\sqrt{(\xi_{mnk} h)^2+(n\pi)^2}}.
\end{equation}
We consider two inertial modes satisfying $n_2-n_1=1$ and $\xi_1\simeq\xi_2$ (this is not a strict constraint on a triadic resonance, but this selection always has a significantly larger growth rate), then we have
\begin{equation}
\frac{|\omega_2|}{|\omega_1|}=\frac{n_2}{n_1}\sqrt{\frac{(\xi_{1} h)^2+(n_1\pi)^2}{(\xi_{2} h)^2+(n_2\pi)^2}}.
\end{equation}
Using $n_2=n_1+1$ and $\xi_1\simeq\xi_2$, we get the inequality
\begin{equation}
\frac{n_2}{n_1}\sqrt{\frac{(\xi_{1} h)^2+(n_1\pi)^2}{(\xi_{2} h)^2+(n_2\pi)^2}}\lesssim \frac{n_2}{n_1}\leqslant 2.
\end{equation}  
On the other hand,
\begin{equation}
\frac{|\omega_2|}{|\omega_1|}=\frac{n_2}{n_1}\sqrt{\frac{(\xi_{1} h)^2+(n_1\pi)^2}{(\xi_{2} h)^2+(n_2\pi)^2}}=\frac{n_2}{n_1}\frac{n_1 \pi}{n_2 \pi}\sqrt{\frac{(\xi_{1} h/n_1\pi)^2+1}{(\xi_{2} h/n_2 \pi)^2+1}}=\sqrt{\frac{(\xi_{1} h/n_1\pi)^2+1}{(\xi_{2} h/n_2 \pi)^2+1}}.
\end{equation}
Again using $n_2=n_1+1$ and $\xi_1\simeq\xi_2$, we have
\begin{equation}
\sqrt{\frac{(\xi_{1} h/n_1\pi)^2+1}{(\xi_{2} h/n_2 \pi)^2+1}}\gtrsim 1.
\end{equation}
So we must satisfy the constraint
\begin{equation}
1\lesssim \frac{|\omega_2|}{|\omega_1|}\lesssim2
\end{equation} 
and $|\omega_1|\leqslant 2$, $|\omega_2|\leqslant 2$.
In order to satisfy $\omega_2-\omega_1=1.0$, there are three possibilities, $|\omega_1|+|\omega_2|=1$,  $|\omega_1|-|\omega_2|=1$ or  $|\omega_2|-|\omega_1|=1$.\\
For the case $|\omega_1|-|\omega_2|=1$, we have
\begin{equation}
\frac{|\omega_2|}{|\omega_1|}=\frac{|\omega_2|}{|\omega_2|+1}<1.
\end{equation}
For the case $|\omega_2|-|\omega_1|=1$, we have
\begin{equation}
\frac{|\omega_2|}{|\omega_1|}=\frac{|\omega_1|+1}{|\omega_1|}>2.
\end{equation}
So those two possibilities are excluded. The only possibility is $|\omega_1|+|\omega_2|=1$, which means the absolute value of the two frequencies are both smaller than 1.0.



\bibliography{precession}

\end{document}